\documentclass[twocolumn,journal]{IEEEtran}
\IEEEoverridecommandlockouts
\usepackage[hidelinks]{hyperref}
\usepackage{cite}
\usepackage{multirow}
\usepackage{amsmath,amssymb,amsfonts,amsthm}
\usepackage{algorithmic}
\usepackage{graphicx}
  \usepackage{dblfloatfix}
\usepackage{textcomp}
\usepackage{xcolor}

\def\BibTeX{{\rm B\kern-.05em{\sc i\kern-.025em b}\kern-.08em
    T\kern-.1667em\lower.7ex\hbox{E}\kern-.125emX}}

\usepackage{enumitem} 
\newlist{casenv}{enumerate}{4}
\setlist[casenv]{leftmargin=*,align=left,widest={iiii}}
\setlist[casenv,1]{label={\itshape \casename~$\mathcal{C}_{\arabic*}:$},ref=\arabic*}
\setlist[casenv,2]{label={{\itshape\ \casename} \roman*:},ref=\roman*}
\setlist[casenv,3]{label={{\itshape\ \casename\ \alph*:}},ref=\alph*}
\setlist[casenv,4]{label={{\itshape\ \casename} \arabic*:},ref=\arabic*}
\providecommand{\casename}{Case}

\newtheorem{thm}{Theorem}

\begin{document}
\title{RIS-Assisted Generalized Receive Quadrature Spatial Modulation with Extension to Multicast Communications}
\author{Mohamad H. Dinan, Khatereh Nadali, and Mark F. Flanagan,~\IEEEmembership{Senior~Member,~IEEE}
\thanks{This work was supported by the Irish Research Council under the Consolidator Laureate Award (grant number IRCLA/2017/209) and by Research Ireland under grant number 24/FFP-P/12895. A preliminary version of this work was presented at IEEE GLOBECOM 2023 \cite{dinan2023globecom}.}
\thanks{Mohamad H. Dinan was with the School of Electrical and Electronic Engineering,
University College Dublin, Dublin, Ireland. He is now with Qorvo Inc., Dublin, Ireland (e-mail: mohamadh.dinan@gmail.com).}
\thanks{Khatereh Nadali was with the School of Electrical and Electronic Engineering,
Technological University Dublin, Dublin, Ireland. She is now with Schneider Electric Ireland, Dublin, Ireland (e-mail: khaterehnadali@gmail.com).}
\thanks{Mark F. Flanagan is with the School of Electrical and Electronic Engineering,
University College Dublin, Dublin, Ireland (e-mail: mark.flanagan@ieee.org).}
}

\maketitle

\begin{abstract}
This paper proposes a novel reconfigurable intelligent surface (RIS)-assisted generalized receive quadrature spatial modulation (RIS-GRQSM) scheme to enhance the spectral efficiency (SE) of RIS-aided \textit{quadrature} spatial modulation (QSM) systems. By leveraging the principle of \textit{generalized} spatial modulation (GSM), multiple receive antennas are independently activated for \textit{both} the in-phase and quadrature components of spatial symbols. To fully exploit the potential of RIS, we formulate a max–min optimization problem to adjust the phase shifts of all RIS elements, thereby maximizing the effective signal-to-noise ratios (SNRs) at the activated antennas. Using Lagrange duality, the original high-dimensional non-convex problem is reduced to a tractable problem with a smaller number of real variables, and a closed-form suboptimal solution is also proposed, which achieves near-optimal performance with a sufficiently large RIS. At the receiver, a low-complexity non-coherent energy-based greedy detector (GD) is introduced for efficient symbol detection. We further extend the RIS-GRQSM framework to a multicast communication system, where all users receive identical information with equal SNR levels, and provide a detailed performance analysis of both systems. In particular, we derive the average bit error probability (ABEP) for the proposed RIS-GRQSM and multicast systems under optimal and suboptimal optimization strategies. Numerical results show that RIS-GRQSM significantly improves the SE and error rate performance compared with benchmark schemes, while the multicast extension achieves performance close to benchmark methods at substantially lower complexity.
\end{abstract}


\section{Introduction\label{sec:intro}}

Reconfigurable intelligent surfaces (RISs) and spatial modulation (SM) are two promising beyond-fifth-generation (B5G) technologies that have attracted significant attention from the research community due to
their potential to enhance the energy efficiency (EE) and spectral efficiency (SE) of wireless communication
systems. RISs can be deployed as passive reflectors in wireless networks
and have the ability to alter the phase of their incident signals, thereby
enabling propagation manipulation \cite{di2019smart,basar2019wireless,liu2022path,alghamdi2020intelligent}. SM, on the other hand, uses the
spatial domain to map information bits to the indices of the transmit
or receive antennas \cite{renzo2011sm,wang2014cellular,di2014spatial,yang2015sm,wen2019sm}. Generalized spatial modulation (GSM) \cite{younis2010generalized} and quadrature spatial modulation (QSM) \cite{mesleh2014quadrature} are variants of SM that have been widely investigated, where the SM principle can be applied either at the transmit or receive antenna side. The combination of RIS and SM has the potential to enable future high-capacity and energy-efficient wireless networks.

Specifically, RIS-space-shift keying (RIS-SSK) and RIS-spatial modulation (RIS-SM) systems were proposed in \cite{basar2020reconfigurable} as the two basic models for RIS-aided index modulation (IM) schemes, in which the RIS forms part of the transmitter and the \emph{receive} antennas are utilized to implement IM. An RIS-based joint transmit and receive antenna IM scheme was proposed in \cite{ma2020large} with a view to targeting high SE. However, the results of \cite{ma2020large} showed that the error rate of the transmit SM bits is significantly higher than that of the receive SM bits; this is due to a decrease in the Euclidean distances introduced by the resulting RIS-assisted channels. By utilizing the \emph{quadrature} SM principle, RIS-aided receive quadrature reflecting modulation (RIS-RQRM) was proposed in \cite{yuan2021receive}. RIS-RQRM involves 
partitioning the RIS into two halves; each half targets the real or imaginary part of the signal at the two selected receive
antennas in order to achieve a doubling of the throughput. However, an inherent drawback of this approach is a significant reduction in the signal-to-noise ratio (SNR) at the receiver. This reduction is a consequence of reducing the number of RIS elements available for each targeted antenna.

In \cite{zhang2022gssk} and \cite{albinsaid2021multiple}, generalized space-shift keying (GSSK) and generalized SM (GSM) approaches have respectively been implemented in an RIS-assisted wireless system. In both scenarios, the RIS is partitioned into multiple parts to target \textit{multiple} antennas at the receiver; hence, the throughput can be increased at the expense of a decrease in the SNR at the target antennas. In \cite{dinan2022ris} and \cite{dinan2023ris}, novel \emph{quadrature} SM schemes were proposed, namely RIS-assisted receive quadrature space-shift keying (RIS-RQSSK) and RIS-assisted receive quadrature spatial modulation (RIS-RQSM), to mitigate the problem of the SNR decrease due to the grouping of the RIS elements. In both scenarios, optimization problems were defined to maximize the SNR of the real part of the signal at one antenna and, at the same time, of the imaginary part of the signal at the second antenna. It was shown that the SE of these approaches increases without any degradation in the SNR.

Although the use of QSM systems enhances the SE, it is desirable to further increase the SE of these systems. However, the options for increasing the SE are limited. One approach is to increase the number of receive antennas; however, this is often not viable in practice. Another option is to employ higher modulation orders. However, numerical results indicate that the system fails to provide reasonable performance when the number of RIS elements is limited, resulting in the emergence of an error floor.

Against this background, in this paper we propose a novel RIS-aided IM scheme, namely RIS-assisted generalized receive quadrature spatial modulation (RIS-GRQSM), which generalizes the system designs of \cite{dinan2022ris} and \cite{dinan2023ris} to provide an enhanced SE. The key contributions of this paper are as follows:
\begin{itemize}
    \item We propose a novel RIS-aided generalized spatial modulation system in which \textit{multiple} antennas are selected independently for the in-phase and quadrature branches of the spatial symbols. The proposed transmitter aims to adjust the phase shifts of the RIS elements to simultaneously maximize the relevant SNR components at \emph{all} of the selected receive antennas. At the receiver, we introduce a highly efficient, low-complexity, non-coherent energy-based greedy detector that directly demodulates the transmitted data bits.
    \item For the RIS phase shift optimization, we define a max-min optimization problem and provide an efficient solution, utilizing Lagrange duality. It is shown that the non-convex optimization problem involving a large number of complex variables is reduced to a system of nonlinear equations with a significantly smaller number of real variables, which can be solved numerically to obtain the optimal solution. Moreover, a low-complexity suboptimal approach is also proposed which can provide a closed-form solution. It is shown that the suboptimal approach provides near-optimal performance in the case where a sufficiently large number of RIS elements is employed.
    \item In addition, we propose a multicast communication scheme that is built on the principle of the RIS-GRQSM scheme. In this scenario, the system operates similarly to the zero-forcing (ZF) precoding technique commonly utilized in multiple-input multiple-output (MIMO) systems, i.e., all users receive identical information with the same level of SNR.
    \item We analyze the average bit error probability (ABEP) of the proposed RIS-GRQSM and multicast schemes under both suboptimal and optimal optimization approaches. It is worth noting that such an analysis is rarely feasible in existing works on RIS-assisted wireless communications, owing to the complexity of the numerical algorithms required for adjusting the RIS phase shifts.
    \item Finally, we evaluate the bit error rate (BER) performance of the systems through numerical simulations. The results demonstrate that the proposed RIS-GRQSM system achieves a notable improvement in SE and significantly outperforms the relevant benchmark schemes. Furthermore, the proposed multicast system attains performance close to that of the benchmark scheme, while offering a substantially lower complexity.
\end{itemize}
The rest of this paper is organized as follows. The system model and transceiver structure of the proposed RIS-GRQSM system are described in Section~\ref{sec:System-Model6}. We provide the problem formulation and its optimal and suboptimal solutions in Section~\ref{sec:optimization6}. The ABEP performance of the proposed
RIS-GRQSM system is analyzed in Section~\ref{sec:performane_analysis}. Section~\ref{sec:Multicast-Communications} presents the proposed multicast communication transceiver design and its ABEP analysis, which can be regarded as a special case of the RIS-GRQSM system. The numerical results and comparisons are provided in Section~\ref{sec:Numerical-Results6}. Finally, Section~\ref{sec:Conclusion6} concludes this paper.

\emph{Notation:} Boldface lower-case letters denote column vectors, and boldface upper-case letters denote matrices. $\left(\cdot\right)^{\Re}$ and $\left(\cdot\right)^{\Im}$ denote the real and imaginary components of a calar/vector, respectively. $\lfloor \cdot \rfloor$ denotes the floor operation. $\left(\cdot\right)^{\star}$ represents the optimum value of a scalar/vector variable. The superscript $(\cdot)^{\rm T}$ denotes the transpose. $\mathbb{E}\left\{ \cdot\right\} $ and $\mathbb{V}\left\{ \cdot\right\}$ respectively denote the
expectation and variance operators. $\mathcal{N}\left(\mu,\sigma^{2}\right)$ (resp., $\mathcal{CN}\left(\mu,\sigma^{2}\right)$) represents the normal (resp., complex normal) distribution with mean $\mu$ and variance $\sigma^{2}$.
$\mathbf{u}\odot\mathbf{v}$ represents the element-wise product of two equal-sized vectors $\mathbf{u}$ and $\mathbf{v}$. $\mathbf{0}$ and $\mathbf{I}_n$ denote the zero vector and the identity matrix of size $n\times n$, respectively. The set $\left\{ 1,2,\dots,K\right\}$  is represented by $[K]$. Finally, the set of complex matrices of size $m\times n$ is denoted by $\mathbb{C}^{m\times n}$.

\section{Proposed RIS-GRQSM System\label{sec:System-Model6}}

\begin{figure}[t]
\begin{centering}
\includegraphics[scale=0.2]{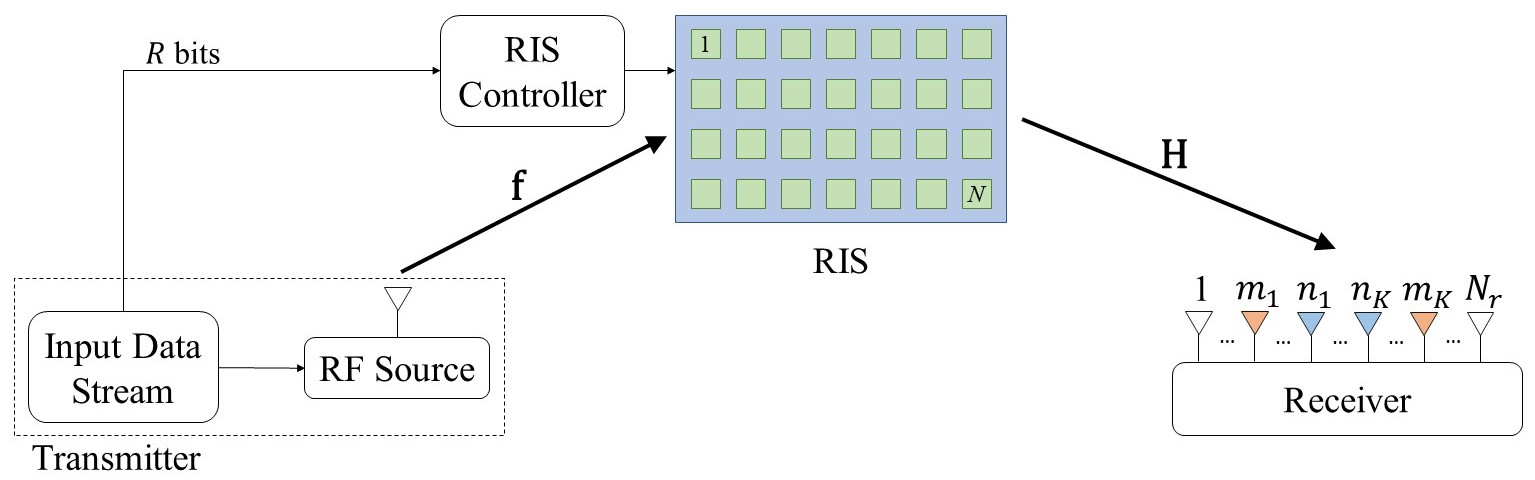}
\par\end{centering}
\caption{A diagrammatic illustration of the RIS-assisted generalized receive quadrature
spatial modulation (RIS-GRQSM) system.\label{fig:Schematic-GRQSM}}
\end{figure}
Fig.~\ref{fig:Schematic-GRQSM} illustrates the transceiver diagram of the proposed RIS-GRQSM system. In the RIS-GRQSM scheme, the transmitter conveys the information bits through the indices of \emph{multiple} selected receive antennas. We assume that the receiver is equipped with $N_{r}$ receive antennas, while the transmitter is equipped with a single antenna that can emit the signal from the RF source toward the RIS. The RIS is located between the receiver and the transmitter and is equipped with $N$ reflecting elements whose phases are adjusted by the transmitter through the RIS controller. We also assume that the direct link between the transmitter and the receiver is blocked by obstacles or severe shadowing. The transmitter selects a group of $K\leq N_{r}$ antennas for the in-phase (I) antenna index modulation, and independently of the first group, it also selects another group of $K$ antennas for the quadrature (Q) index modulation, so that the transmitter is able to transmit $2\times\left\lfloor \log_{2}\binom{N_{r}}{K}\right\rfloor $ bits via IM. In addition, the polarity of the real and imaginary parts of the signals at the selected antennas can be controlled by the transmitter to convey an additional single bit per each selected I or Q antenna index. Therefore, the total rate of the proposed RIS-GRQSM system is $R=2\left(K+\left\lfloor \log_{2}\binom{N_{r}}{K}\right\rfloor \right)$ bits per channel use (bpcu).

In this wireless communication model, the baseband received signal is given by 
\begin{equation}
\mathbf{y}=\sqrt{E_{s}}\mathbf{H}\left(\boldsymbol{\theta}\odot\mathbf{f}\right)+\mathbf{w},\label{eq:received signal}
\end{equation}
where $\mathbf{H}\in\mathbb{C}^{N_{r}\times N}$ and $\mathbf{f}=[\beta_{1}e^{j\psi_{1}},\beta_{2}e^{j\psi_{2}},\dots,\beta_{N}e^{j\psi_{N}}]^{\rm T}\in\mathbb{C}^{N\times1}$
respectively are the fading channels between the receiver and the
RIS, and between the RIS and the single-antenna transmitter, whose
elements are independent and identically distributed (i.i.d.) according
to $\mathcal{CN}(0,1)$, $\boldsymbol{\theta}=[e^{j\phi_{1}},e^{j\phi_{2}},\dots,e^{j\phi_{N}}] ^{\rm T}\in\mathbb{C}^{N\times1}$
is the vector of reflection coefficients of the RIS elements, $\mathbf{w}=[w_{1},w_{2},\dots,w_{N_{r}}] ^{\rm T}\in\mathbb{C}^{N_{r}\times1}$
is the vector of additive white Gaussian noise at the $N_{r}$ receive
antennas which is distributed according to $\mathcal{CN}(\mathbf{0},N_{0}\mathbf{I}_{N_{r}})$,
and $E_{s}$ is the transmitted energy per symbol from the RF source. Hence the SNR is $E_s/N_0$.

In the proposed RIS-GRQSM system, the transmitter aims to simultaneously
maximize the SNR of the I and Q components of the selected receive
antennas in order that these antenna activations can be easily detected by a simple energy-based greedy
detector (GD). Using (\ref{eq:received signal}), we may write expressions for the real (in-phase) and imaginary (quadrature) components of the
received signal at the selected antennas, respectively, as 
\begin{align}
y_{m_{k}}^{\Re} =& \sqrt{E_{s}}\left[\mathbf{h}_{m_{k}}^{\Re}\left(\boldsymbol{\theta}\odot\mathbf{f}\right)^{\Re}-\mathbf{h}_{m_{k}}^{\Im}\left(\boldsymbol{\theta}\odot\mathbf{f}\right)^{\Im}\right]+w_{m_{k}}^{\Re} \nonumber\\
 =& \sqrt{E_{s}}\sum_{i=1}^{N}\beta_{i}\left(h_{m_{k},i}^{\Re}\omega_{i}^{\Re}-h_{m_{k},i}^{\Im}\omega_{i}^{\Im}\right)+w_{m_{k}}^{\Re},\;k \in [K], \label{eq:y_real}
\end{align}
and
\begin{align}
y_{n_{k}}^{\Im} =& \sqrt{E_{s}}\left[\mathbf{h}_{n_{k}}^{\Re}\left(\boldsymbol{\theta}\odot\mathbf{f}\right)^{\Im}+\mathbf{h}_{n_{k}}^{\Im}\left(\boldsymbol{\theta}\odot\mathbf{f}\right)^{\Re}\right]+w_{n_{k}}^{\Im} \nonumber\\
 =& \sqrt{E_{s}}\sum_{i=1}^{N}\beta_{i}\left(h_{n_{k},i}^{\Re}\omega_{i}^{\Im}+h_{n_{k},i}^{\Im}\omega_{i}^{\Re}\right)+w_{n_{k}}^{\Im},\;k \in [K], \label{eq:y_imag}
\end{align}
where $\{ m_1, m_2, \ldots, m_K \} \triangleq \mathcal{M} \subseteq [N_r]$ and $\{ n_1, n_2, \ldots, n_K \} \triangleq \mathcal{N} \subseteq [N_r]$ are the sets of selected antennas for the I and Q components, respectively, $\mathbf{h}_{l}=[h_{l,1},h_{l,2},\dots,h_{l,N}]$
is the $l$-th row of $\mathbf{H}$, and we define $\omega_{i}=e^{j(\psi_{i}+\phi_{i})}$.

\subsection*{Receiver structure}

At the receiver, a successive greedy detector (GD) can be employed
to detect the selected receive antennas. The GD operates via 
\begin{equation}
\hat{\mathcal{M}}=\{ \hat{m}_k \}_{k \in [K]}=\arg {\max}_{K}\left\{ |y_{m}^{\Re}|\right\}_{m\in[N_{r}]} , \label{eq:M_hat}
\end{equation}
\begin{equation}
\hat{\mathcal{N}}=\{\hat{n}_{k}\}_{k\in[K]}=\arg{\max}_{K}\left\{ |y_{n}^{\Im}|\right\}_{n\in[N_{r}]} , \label{eq:N_hat}
\end{equation}
where $\arg{\max}_{K}\{\cdot\}$ finds the arguments of the $K$ largest
values in a set. Then, the receiver maps the set of antennas to the
information bits according to the predefined codebook used at the transmitter.
After detecting the set of selected antennas, the polarity bits
can be detected by testing the polarity of each of the values of $y_{\hat{m}_{k}}^{\Re}$
and $y_{\hat{n}_{k}}^{\Im}$ for each $k\in[K]$.

\section{Problem Formulation and Solution \label{sec:optimization6}}

In this section, we propose an optimization problem to design the phase shifts of the RIS elements and we provide optimal and suboptimal solutions to this problem. In order to maximize the SNR at the I and Q components of the selected
antennas, we define the following max-min optimization problem 
\begin{align}
\underset{\{\omega_{i}^{\Re}\},\{\omega_{i}^{\Im}\}}{\max} & \min\{X_k:k\in[K]\}\cup\{Y_k:k\in[K]\}  \label{eq: max-min-QSM}\\
\mbox{where}\;\;\;\; & X_k=x_{{\rm P},k}^{\rm R}\sum_{i=1}^{N}\beta_{i}\left(h_{m_{k},i}^{\Re}\omega_{i}^{\Re}-h_{m_{k},i}^{\Im}\omega_{i}^{\Im}\right),\; k \in [K],\nonumber \\
 & Y_k=x_{{\rm P},k}^{\rm I}\sum_{i=1}^{N}\beta_{i}\left(h_{n_{k},i}^{\Re}\omega_{i}^{\Im}+h_{n_{k},i}^{\Im}\omega_{i}^{\Re}\right),\; k \in [K], \nonumber \\
\mbox{s.t.}\;\;\;\;\;\; & \left(\omega_{i}^{\Re}\right)^{2}+\left(\omega_{i}^{\Im}\right)^{2}=1,\ \mbox{for all}\;i \in [N],\nonumber 
\end{align}
where $x_{{\rm P},k}^{\rm R}\in\{-1,1\}$ and $x_{{\rm P},k}^{\rm I}\in\{-1,1\}$,
$k \in [K]$, are the polarity symbols associated with the transmitted polarity bit for the real part of antenna $m_k$ and the imaginary part of antenna $n_k$, respectively. We can observe that the optimum values obtained by the problem above must be positive; therefore, the polarity symbols identify the polarity of
the real and imaginary components of the noise-free received signals
at the selected antennas. The optimal solution ($\{ \omega_i^{\Re}, \omega_i^{\Im} \}$ for all $i \in [N]$) can be derived by finding the minimizing point of the Lagrange function
associated with the problem above \cite{boyd2004convex}. A similar approach is used in \cite{dinan2022ris} to solve the problem for the case
where $K=1$. In that case, the minimization involves two functions, while for the case where $K>1$, the minimization involves $2K$ functions associated with the selected antennas; hence, the max-min problem (\ref{eq: max-min-QSM}) can be transformed into a maximization problem with $2K$ linear inequality constraints by defining an auxiliary parameter (cf. \cite[Eq. (10)]{dinan2022ris}). Then, its corresponding Lagrange function can be derived (cf. \cite[Eq. (11)]{dinan2022ris}). Therefore, by applying the optimality conditions on the corresponding Lagrange function, the optimal values of $\omega_{i}^{\Re}$ and $\omega_{i}^{\Im}$
can be derived respectively as (\ref{eq:optimum theta_R6}) and (\ref{eq:optimum theta_I6}), shown at the top of the next page,
\begin{figure*}[t]
\begin{equation}
\omega_{i}^{\Re\star}=\frac{\sum_{k=1}^{K}\lambda_{k}A_{k,i}+\sum_{k=1}^{K}\delta_{k}B_{k,i}}{\sqrt{\left(\sum_{k=1}^{K}\lambda_{k}A_{k,i}+\sum_{k=1}^{K}\delta_{k}B_{k,i}\right)^{2}+\left(\sum_{k=1}^{K}\lambda_{k}C_{k,i}+\sum_{k=1}^{K}\delta_{k}D_{k,i}\right)^{2}}},\label{eq:optimum theta_R6}
\end{equation}

\begin{equation}
\omega_{i}^{\Im\star}=\frac{\sum_{k=1}^{K}\lambda_{k}C_{k,i}+\sum_{k=1}^{K}\delta_{k}D_{k,i}}{\sqrt{\left(\sum_{k=1}^{K}\lambda_{k}A_{k,i}+\sum_{k=1}^{K}\delta_{k}B_{k,i}\right)^{2}+\left(\sum_{k=1}^{K}\lambda_{k}C_{k,i}+\sum_{k=1}^{K}\delta_{k}D_{k,i}\right)^{2}}}.\label{eq:optimum theta_I6}
\end{equation}
\rule{\linewidth}{0.4pt}
\end{figure*}
for all $i \in [N]$, where $\lambda_{k}\geq0$ and $\delta_{k}\geq0$,
$k \in [K]$, are the Lagrange multipliers, and to simplify the
notation, we define $A_{k,i}\triangleq x_{{\rm P},k}^{\rm R}h_{m_{k},i}^{\Re}$,
$B_{k,i}\triangleq x_{{\rm P},k}^{\rm I}h_{n_{k},i}^{\Im}$, $C_{k,i}\triangleq -x_{{\rm P},k}^{\rm R}h_{m_{k},i}^{\Im}$,
and $D_{k,i}\triangleq x_{{\rm P},k}^{\rm I}h_{n_{k},i}^{\Re}$.

Next, the Lagrange dual problem can be defined as (\ref{eq:convex problem6}), shown at the top of the next page. 
\begin{figure*}[t]
\begin{align}
\min_{\{\lambda_{k}\},\{\delta_{k}\}} & \;\sum_{i=1}^{N}\beta_{i}\sqrt{\left(\sum_{k=1}^{K}\lambda_{k}A_{k,i}+\sum_{k=1}^{K}\delta_{k}B_{k,i}\right)^{2}+\left(\sum_{k=1}^{K}\lambda_{k}C_{k,i}+\sum_{k=1}^{K}\delta_{k}D_{k,i}\right)^{2}}\label{eq:convex problem6}\\
\text{s.t.}\;\; & \;\sum_{k=1}^{K}\lambda_{k}+\sum_{k=1}^{K}\delta_{k}=1,\quad \mbox{and}\quad \lambda_{k},\delta_{k}\geq0,\quad k \in [K].\nonumber 
\end{align}
\rule{\linewidth}{0.4pt}
\end{figure*}
This problem is convex and can be solved numerically using known convex optimization methods. It is worth noting that the non-convex optimization
problem in (\ref{eq: max-min-QSM}) with $N\gg1$ complex variables
is reduced to a convex problem with $2K\ll N$ real variables. Another approach is to utilize the Karush-Kuhn-Tucker (KKT) conditions \cite{boyd2004convex} to transform this convex optimization
into the system of $2K$ equations given by (\ref{eq:sys_eq_1}) to (\ref{eq:sys_eq_3}), shown at the top of the next page,
\begin{figure*}[t]
\begin{align}
 & \sum_{i=1}^{N}\beta_{i}\frac{\left(\sum_{k=1}^{K}\lambda_{k}A_{k,i}+\sum_{k=1}^{K}\delta_{k}B_{k,i}\right)\left(A_{1,i}-A_{k',i}\right)+\left(\sum_{k=1}^{K}\lambda_{k}C_{k,i}+\sum_{k=1}^{K}\delta_{k}D_{k,i}\right)\left(C_{1,i}-C_{k',i}\right)}{\sqrt{\left(\sum_{k=1}^{K}\lambda_{k}A_{k,i}+\sum_{k=1}^{K}\delta_{k}B_{k,i}\right)^{2}+\left(\sum_{k=1}^{K}\lambda_{k}C_{k,i}+\sum_{k=1}^{K}\delta_{k}D_{k,i}\right)^{2}}}=0,k'=2,\dots,K,\label{eq:sys_eq_1}\\
 & \sum_{i=1}^{N}\beta_{i}\frac{\left(\sum_{k=1}^{K}\lambda_{k}A_{k,i}+\sum_{k=1}^{K}\delta_{k}B_{k,i}\right)\left(A_{1,i}-B_{k',i}\right)+\left(\sum_{k=1}^{K}\lambda_{k}C_{k,i}+\sum_{k=1}^{K}\delta_{k}D_{k,i}\right)\left(C_{1,i}-D_{k',i}\right)}{\sqrt{\left(\sum_{k=1}^{K}\lambda_{k}A_{k,i}+\sum_{k=1}^{K}\delta_{k}B_{k,i}\right)^{2}+\left(\sum_{k=1}^{K}\lambda_{k}C_{k,i}+\sum_{k=1}^{K}\delta_{k}D_{k,i}\right)^{2}}}=0,k'=1,\dots,K,\label{eq:sys_eq_2}\\
 & \sum_{k=1}^{K}\lambda_{k}+\sum_{k=1}^{K}\delta_{k}=1,\label{eq:sys_eq_3}
\end{align}
\rule{\linewidth}{0.4pt}
\end{figure*}
where the first $2K-1$ equations are nonlinear. This system of $2K$
equations with $2K$ variables can be solved numerically to obtain the optimal values of 
$\{\lambda_{k}\}$ and $\{\delta_{k}\}$.

\subsection*{Suboptimal approach}

Next, we will show how a simplified optimization procedure can be used to provide a suboptimal and practical solution to this problem in a way that avoids the need to solve (\ref{eq:convex problem6}) exactly. Note that the solution for a specific
variable $\lambda_{k}$ in (\ref{eq:convex problem6}) is a function
of all variables $\{\beta_{i}\}$, $\{A_{k,i}\}$, $\{B_{k,i}\}$,
$\{C_{k,i}\}$ and $\{D_{k,i}\}$, i.e., $\lambda_{k}=f\left(\{\beta_{i}\}_{i\in[N]},\gamma(\mathcal{V})\right)$,
where $\gamma(\mathcal{V})$ is a specific permutation of the set
of variables 
\begin{align*}
    \mathcal{V}=& \{A_{k,i}\}_{k\in[K],i\in[N]} \cup \{B_{k,i}\}_{k\in[K],i\in[N]} \\
    & \cup \{C_{k,i}\}_{k\in[K],i\in[N]} \cup \{D_{k,i}\}_{k\in[K],i\in[N]} .
\end{align*}
From the problem (\ref{eq:convex problem6}), it can be seen that there is symmetry between each pair of variables in the set
$ \{\lambda_{k}\}_{k\in[K]}\cup \{\delta_{k}\}_{k\in[K]} $.
Hence, it can be concluded that the solution for any $\lambda_{k'}$ ($k'\neq k$)
or $\delta_{k}$ is given by $\lambda_{k'}=f\left(\{\beta_i\}_{i\in[N]},\gamma'(\mathcal{V})\right)$ or $\delta_{k}=f\left(\{\beta_i\}_{i\in[N]},\gamma''(\mathcal{V})\right)$, where $\gamma'(\mathcal{V})$ or $\gamma''(\mathcal{V})$ are permutations of $\mathcal{V}$ different from $\gamma(\mathcal{V})$.
We know that all variables in $\mathcal{V}$ are identically distributed
according to $\mathcal{N}(0,1/2)$; as a result, $\lambda_{k}$ and $\delta_{k}$, $k \in [K]$, are random variables (RVs) with identical mean values,
that is, $\mathbb{E}\{\lambda_{k}\}=\mathbb{E}\{\delta_{k}\}=1/(2K)$ (note that $\sum_{k}\lambda_{k}+\sum_{k}\delta_{k}=1$). 
On the other hand, experimental results show that as the number of RIS elements $N$ increases, the variances of all of the variables $\lambda_k$ and $\delta_k$ tend to zero. Hence, for sufficiently large values of $N$ ($N\gg2K$), a suboptimal solution can be found by simply approximating each of $\lambda_k$ and $\delta_k$ by their mean values, i.e., 
\begin{equation}
    \lambda_k=\delta_k=\frac{1}{2K}, \forall k\in[K]. \label{eq:approximateLambda}
\end{equation}
To gain further insights, in Table~\ref{tab:lambda_mean_var} we present numerical results for the mean and variance of $\lambda_1$, as an example, for the case where $K=2$ and for different values of $N$ (here $10^4$ channel realizations were used). It can be observed that with a sufficiently large value of $N$, $\mathbb{E}\{\lambda_{1}\}$ becomes very close to $1/(2K)=1/4$, and its variance tends to zero with an increasing number of RIS elements.

\begin{table}[t]
\caption{\textsc{Mean and variance of $\lambda_{1}$ for the case where $K=2$ and for different
values of $N$.}\label{tab:lambda_mean_var}}

\centering{}%
\begin{tabular}{|c|c|c|c|}
\hline 
 & $N=128$ & $N=256$ & $N=512$\tabularnewline
\hline 
\hline 
Mean & $0.2490$ & $0.2502$ & $0.2500$\tabularnewline
\hline 
Variance & $0.0011$ & $5.25\times10^{-4}$ & $2.6110\times10^{-4}$\tabularnewline
\hline 
\end{tabular}
\end{table}

{\color{black}
\section{Performance Analysis\label{sec:performane_analysis} of RIS-GRQSM system}
In this section, we examine the ABEP of the proposed RIS-GRQSM system, with a focus on the GD receiver. We begin by considering a suboptimal approach, as it enables a closed-form expression for determining the optimal phase shifts of the RIS elements. This closed-form solution allows for an analytical evaluation of the ABEP performance. Additionally, it provides valuable insight into assessing the system’s performance when optimal phase shifts are applied. Note that this is in marked contrast to the vast majority of works on RIS-assisted wireless communications, where such an analysis is not possible due to the fact that the RIS phase shifts are optimized using a complex algorithm whose output cannot be characterized analytically.

\subsection{Analysis based on the suboptimal approach}
In this subsection, we analyze the ABEP of the system in which the RIS phase shifts are determined using (\ref{eq:optimum theta_R6}) and (\ref{eq:optimum theta_I6}), where the simplifying expression in (\ref{eq:approximateLambda}) is used to obtain suboptimal phase shifts. In the following, the ABEP analysis is performed only for the detection of the spatial symbol $\mathcal{M}$ which identifies the antennas with active real parts along with the corresponding polarity bits; due to the inherent symmetry in the expressions, it is easy to show that the ABEP expression for the detection of the spatial symbol $\mathcal{N}$ and its corresponding polarity bits is identical. An upper bound on the ABEP, which is tight at high SNR, is given by 
\begin{align}
{\rm ABEP}_{{\rm ub}} = & \frac{1-{\rm P}_{e}\left(\mathcal{M}\right)}{K+\left\lfloor \log_{2}\binom{N_{r}}{K}\right\rfloor }\sum_{k=1}^{K}k\binom{K}{k} \nonumber \\
& \times {\rm Pr}\left(\mathcal{E} = \mathcal{E}_k|\hat{\mathcal{M}}=\mathcal{M}\right)+ \rho{\rm P}_{e}\left(\mathcal{M}\right),\label{eq:ABEP}
\end{align}
where ${\rm P}_{e}\left(\mathcal{M}\right)$ is the probability of erroneous detection of $\mathcal{M}$, i.e., of the set of the selected receive antennas with active real part, $\rho$ is the average fraction of bits in error when $\mathcal{M}$ is wrongly detected, which in the worst-case scenario is equal to $1/2$, and ${\rm Pr}\big(\mathcal{E} = \mathcal{E}_k|\hat{\mathcal{M}}=\mathcal{M}\big)$ is the probability of erroneous detection of a particular set $\mathcal{E}_k$ of $k$ polarity symbols conditioned on correct detection of the corresponding antenna indices (here $\mathcal{E}$ denotes the set of erroneously detected polarity symbols).

Next, ${\rm P}_{e}\left(\mathcal{M}\right)$ is given by 
\begin{align}
{\rm P}_{e}\left(\mathcal{M}\right)=&\sum_{\hat{\mathcal{M}}\neq\mathcal{M}}{\rm PEP}\left(\mathcal{M}\rightarrow\hat{\mathcal{M}}\right)\nonumber \\
\approx & \sum_{\substack{\hat{\mathcal{M}}\neq\mathcal{M}, \\ \left|\hat{\mathcal{M}}\backslash\mathcal{M}\right|=1}} {\rm PEP}\left(\mathcal{M}\rightarrow\hat{\mathcal{M}}\right),
\end{align}
where ${\rm PEP}\left(\mathcal{M}\rightarrow\hat{\mathcal{M}}\right)$ is the pairwise error probability (PEP) associated with the spatial symbols $\mathcal{M}$ and $\hat{\mathcal{M}}$. Here we consider the fact that at high SNR values, ${\rm P}_{e}\left(\mathcal{M}\right)$ is dominated by error sets ${\hat{\mathcal{M}}}$ that differ from ${\mathcal{M}}$ by a single element, i.e., the index of a single antenna is erroneously detected. Hence, ${\rm P}_{e}\left(\mathcal{M}\right)$ can be expressed as 
\begin{align}
{\rm P}_{e}\left(\mathcal{M}\right) \approx &\sum_{k=1}^K \sum_{q=K+1}^{N_r} {\rm Pr}\big(\big|y_{m_q}^{\Re}\big|>\big|y_{m_k}^{\Re}\big| \ |m_k\in\mathcal{M},m_q\notin\mathcal{M} \big) \nonumber\\
= & K(N_r-K) {\rm Pr}\big(\big|y_{m_q}^{\Re}\big|>\big|y_{m_k}^{\Re}\big|\ | m_k\in\mathcal{M}, m_q\notin\mathcal{M}\big).
\end{align}
In addition, by a similar reasoning, we can approximate the error probability corresponding to the polarity bits by the error probability of a single bit, i.e., the error caused by the wrong detection of a single \textit{polarity} symbol, since ${\rm Pr}(\mathcal{E}=\mathcal{E}_k | \hat{\mathcal{M}}=\mathcal{M}) \approx \big[{\rm Pr}(\mathcal{E}=\mathcal{E}_1 | \hat{\mathcal{M}}=\mathcal{M} ) \big]^k \ll {\rm Pr}(\mathcal{E} = \mathcal{E}_1 | \hat{\mathcal{M}}=\mathcal{M})$ for $k>1$, since for large values of $N$ the error is mainly dependent on the noise, which has the effect of making these errors approximately independent. Therefore,  the $k=1$ term in the summation in (\ref{eq:ABEP}) dominates. Hence, an approximate upper bound on the ABEP can be expressed as 
\begin{align}
& {\rm ABEP}_{\rm ub} \nonumber \\
& \approx \frac{1-{\rm P}_{e}\left(\mathcal{M}\right)}{K+\left\lfloor \log_{2}\binom{N_{r}}{K}\right\rfloor }K\ {\rm Pr}\left(\mathcal{E}=\mathcal{E}_1 | \hat{\mathcal{M}}=\mathcal{M}\right)\nonumber \\
  & \quad +\rho K(N_r-K){\rm Pr} \big(\big|y_{m_q}^{\Re}\big|>\big|y_{m_k}^{\Re}\big|\ | m_k\in\mathcal{M}, m_q\notin\mathcal{M}\big).\label{eq:ABEP_2}
\end{align}

\begin{figure*}[!b]
{\color{black}
\rule{\linewidth}{0.4pt}
\begin{equation}
    y_{m_\ell}^{\Re}= \sqrt{E_s}\sum_{i=1}^N \beta_i \frac{A_{\ell,i}\sum_{j=1}^{K}A_{j,i}+A_{\ell,i}\sum_{j=1}^{K}B_{j,i}+C_{\ell,i}\sum_{j=1}^{K}C_{j,i}+C_{\ell,i}\sum_{j=1}^{K}D_{j,i}}{\sqrt{\left(\sum_{j=1}^{K}A_{j,i}+\sum_{j=1}^{K}B_{j,i}\right)^{2}+\left(\sum_{j=1}^{K}C_{j,i}+\sum_{j=1}^{K}D_{j,i}\right)^{2}}}+w_{m_\ell}^{\Re},\ \ell\in\{k\in[K],q\notin[K]\} \label{eq:Z}
\end{equation}
}
\end{figure*}

To evaluate (\ref{eq:ABEP_2}), the probability density functions (PDFs) of $y_{m_k}^{\Re}$ and $y_{m_q}^\Re$ are required. Substituting (\ref{eq:optimum theta_R6}) and (\ref{eq:optimum theta_I6}) into (\ref{eq:y_real}) and considering $\lambda_k=\delta_k=1/(2K)$ for all $k\in[K]$, the received signal at antenna $m_\ell$ is given by (\ref{eq:Z}), shown at the bottom of the next page, where $\ell\in\{k,q\}$. Here, we consider positive polarity symbols for both the real and imaginary parts; it is clear that the same analysis applies to the other cases. From this point onward, We define $Z\triangleq y_{m_k}^\Re=\sqrt{E_s}\sum_{i=1}^N Z_i+w_{m_k}^{\Re}$ and $\hat{Z}\triangleq y_{m_q}^\Re=\sqrt{E_s}\sum_{i=1}^N \hat{Z}_i+w_{m_q}^{\Re}$ to simplify the notation. According to the central limit theorem (CLT), with a large number of RIS elements, the PDFs of $Z$ and $\hat{Z}$ are both very well approximated by normal distributions, the means and variances of which can be derived by calculating the means and variances of the terms $Z_i$ and $\hat{Z}_i$. These calculations are not straightforward, since the selected antennas $\{n_k\}$ associated with the imaginary parts of the IM symbol also contribute to $Z$ and $\hat{Z}$. Hence, the following scenarios must be examined:
\begin{casenv}
\item $m_k,m_q\notin \mathcal{N}$;
\vspace{0.2cm}
\item $m_k\notin \mathcal N\ {\rm and}\ m_q\in \mathcal{N}$;
\vspace{0.2cm}
\item $m_k\in \mathcal{N} \ {\rm and}\ m_q\notin\mathcal{N}$;
\vspace{0.2cm}
\item $m_k,m_q\in\mathcal{N}$.
\end{casenv}
\begin{thm} \label{thm:mean_V}
    In each of the cases above, $Z$ and $\hat{Z}$ follow normal distributions, with respective means and variances given by 
    \begin{casenv}
        \item $$\mu_1=\frac{N\pi\sqrt{E_s}}{4\sqrt{2K}},\ \sigma_1^2=\left(\frac{N}{2}+\frac{8-\pi^2}{32}\frac{N}{K}\right)E_s+\frac{N_0}{2};$$
        $$\hat{\mu}_1=0,\ \hat{\sigma}_1^2=\frac{N}{2}E_s+\frac{N_0}{2};$$
        \item $$\mu_2=\mu_1,\ \sigma_2=\sigma_1;$$
        $$\hat{\mu}_2=0,\ \hat{\sigma}_2^2=\left(\frac{N}{2}-\frac{1}{4}\frac{N}{K}\right)E_s+\frac{N_0}{2};$$
        \item $$\mu_3=\mu_1,\ \sigma_3^2=\left(\frac{N}{2}-\frac{\pi^2}{32}\frac{N}{K}\right)E_s+\frac{N_0}{2};$$
        $$\hat{\mu}_3=\hat{\mu}_1,\ \hat{\sigma}_3=\hat{\sigma}_1;$$
        \item $$\mu_4=\mu_3,\ \sigma_4=\sigma_3;$$
        $$\hat{\mu}_4=\hat{\mu}_2,\ \hat{\sigma}_4=\hat{\sigma}_2.$$
    \end{casenv}
\end{thm}
\begin{proof} The proof is provided in Appendix \ref{app:appendix1}.
\end{proof}

Moreover, note that the probability of occurrence of each case is given by\footnote{It is worth mentioning that these probabilities are approximate, since not all combinations of $\binom{N_r}{K}$ are utilized in antenna selection; nevertheless, they provide a close estimate of the actual probabilities.}
\begin{casenv}
    \item $$P_1 = \sum_{k=0}^K \frac{\binom{K}{k}\binom{N_r-K}{K-k}}{\binom{N_r}{K}} \times \frac{K-k}{K} \times \frac{N_r-2K+k}{N_r-K}$$
    \item $$P_2 = \sum_{k=0}^K \frac{\binom{K}{k}\binom{N_r-K}{K-k}}{\binom{N_r}{K}} \times \frac{K-k}{K} \times \frac{K-k}{N_r-K}$$
    \item $$P_3 = \sum_{k=0}^K \frac{\binom{K}{k}\binom{N_r-K}{K-k}}{\binom{N_r}{K}} \times \frac{k}{K} \times \frac{N_r-2K+k}{N_r-K}$$
    \item $$P_4 = \sum_{k=0}^K \frac{\binom{K}{k}\binom{N_r-K}{K-k}}{\binom{N_r}{K}} \times \frac{k}{K} \times \frac{K-k}{N_r-K}.$$
\end{casenv}
The expressions above represent the probabilities associated with the four possible cases. Specifically:
\begin{enumerate}[label=(\roman*)]
    \item The first factor in each term represents the probability that the two antenna sets, $\mathcal{M}$ and $\mathcal{N}$, share exactly $k$ common antennas.
    \item Conditioned on this event, the second factor corresponds to the probability that $m_k$ either belongs to or does not belong to $\mathcal{N}$, corresponding to \textit{Cases}~$\mathcal{C}_1$ and $\mathcal{C}_2$, and \textit{Cases}~$\mathcal{C}_3$ and $\mathcal{C}_4$, respectively.
    \item Finally, the third factor represents the probability that $m_q$ is excluded from (respectively, included in) $\mathcal{N}$ for \textit{Cases}~$\mathcal{C}_1$ and $\mathcal{C}_3$ (respectively, \textit{Cases}~$\mathcal{C}_2$ and $\mathcal{C}_4$), conditioned on the first two events.
\end{enumerate}
Then, we can write 
\begin{align}
    & {\rm Pr}\big\{|Z|<|\hat{Z}|\big\} \nonumber\\
    & = \sum_{c=1}^{4} P_c \Big[ \int_0^{\infty} p_{Z|\mathcal{C}_c}(\alpha) {\rm Pr}\left\{|\hat{Z}| > \alpha | \mathcal{C}_c\right\} d\alpha  \nonumber \\
    & \quad +  \int_0^{\infty} p_{Z|\mathcal{C}_c}(-\alpha) {\rm Pr}\left\{|\hat{Z}| > \alpha | \mathcal{C}_c\right\} d\alpha \Big] \nonumber \\
    & = 2\sum_{c=1}^4 P_c\Big[ \int_0^\infty p_{Z|\mathcal{C}_c}(\alpha){\rm Pr}\left\{\hat{Z}>\alpha\ |\mathcal{C}_c\right\}{\rm d}\alpha \nonumber \\
    & \quad + \int_0^\infty p_{Z|\mathcal{C}_c}(-\alpha) {\rm Pr}\left\{\hat{Z}>\alpha \ |\mathcal{C}_c\right\}{\rm d}\alpha\Big], \label{eq:Pr1}
\end{align}
where $p_{Z|\mathcal{C}_c}(\alpha)$ is the PDF of $Z$ conditioned on case $\mathcal{C}_c$. Using Theorem~\ref{thm:mean_V}, the above probability can be expressed in the form (\ref{eq:Pr2}), shown at the bottom of the next page.

\begin{figure*}[!b]
{\color{black}
\rule{\linewidth}{0.4pt}
\begin{align}
    {\rm Pr}\left\{|Z|<|\hat{Z}|\right\} = & \sum_{c=1}^4 P_c\frac{\sqrt{2}}{\sqrt{\pi}\sigma_c}\Bigg[\int_0^\infty e^{\frac{-(\mu_c-\alpha)^2}{2\sigma_c^2}} {\rm Q}\left(\frac{\alpha}{\hat{\sigma}_c}\right){\rm d}\alpha+
     \int_0^\infty e^{\frac{-(\mu_c+\alpha)^2}{2\sigma_c^2}} {\rm Q}\left(\frac{\alpha}{\hat{\sigma}_c}\right){\rm d}\alpha\Bigg]. \label{eq:Pr2}
\end{align}
}
\end{figure*}

Next, the probability corresponding to the polarity symbols, i.e., ${\rm Pr}\big\{\mathcal{E}=\mathcal{E}_1|\hat{\mathcal{M}}=\mathcal{M}\big\}$, is derived as 
\begin{align}
    & {\rm Pr}\big\{\mathcal{E}=\mathcal{E}_1|\hat{\mathcal{M}}=\mathcal{M}\big\} \nonumber \\
    & =(P_3+P_4)\mathbb{E}_X\Bigg\{{\rm Q}\Bigg(\sqrt{X^2 \frac{2E_s}{N_0}}\Bigg)|m_k \in\mathcal{N}\Bigg\} \nonumber \\
    & \quad +(1-P_3-P_4)\mathbb{E}_X\Bigg\{{\rm Q}\Bigg(\sqrt{X^2 \frac{2E_s}{N_0}}\Bigg)|m_k\notin\mathcal{N}\Bigg\}, \label{eq:PrPol}
\end{align}
where $X\triangleq \sum_i Z_i$, and, as shown in Theorem \ref{thm:mean_V}, the variance of $X$ is evaluated depending on whether $m_k\in\mathcal{N}$ (corresponding to \textit{Cases}~$\mathcal{C}_3$ and $\mathcal{C}_4$) or not (corresponding to \textit{Cases}~$\mathcal{C}_1$ and $\mathcal{C}_2$), while $\mathbb{E}_V\big\{{\rm Q}\big(\sqrt{V}\big)\big\}$ is given by \cite{craig1991new}
\begin{equation}
    \mathbb{E}_V\big\{{\rm Q}\big(\sqrt{V}\big)\big\}=\frac{1}{\pi}\int_0^{\pi/2} {\rm M}_V\bigg(\frac{-1}{2\sin^2\Phi}\bigg){\rm d}\Phi, \label{eq:craig}
\end{equation}
where ${\rm M}_V(t)$ is the moment generating function (MGF) of RV $V$.

Finally, substituting (\ref{eq:Pr2}) and (\ref{eq:PrPol}) into (\ref{eq:ABEP_2}), we can derive an approximate upper bound on the ABEP.

\subsection{ABEP of the system with optimal phase shifts}
In this subsection, by leveraging the analysis of the suboptimal approach and taking into account a noteworthy observation, we derive the approximate ABEP of the proposed RIS-GRQSM system with optimal phase shifts. Referring to Theorem~\ref{thm:mean_V}, numerical results strongly suggest that, with optimal phase shifts, the mean and variance of the signal at a selected receive antenna in $\mathcal{M}$ remain unaffected by the chosen antenna set $\mathcal{N}$. Consequently, the probability scenarios can be reduced to two distinct cases: whether the non-selected antenna $m_q$ belongs to the set $\mathcal{N}$ or not, i.e., i) $m_q\notin \mathcal{N}$, and ii) $m_q\in \mathcal{N}$. In fact, using the optimal values of $\{\lambda_k\}$ and $\{\delta_k\}$ significantly reduces the signal variance at the selected antennas, thereby leading to improved system performance. In addition, numerical simulations show that the variance of the noise-free signal at each of the selected antennas is approximately given by
\begin{equation}
    \sigma^2\approx NE_s\left(\frac{1}{2K}-\frac{\pi^2}{32K}\right). \label{eq:optimal_var}
\end{equation}
Note that the second term in the parentheses in (\ref{eq:optimal_var}) relates to $\mathbb{E}^2\{Z_i\}$, that is equal to the corresponding value in the suboptimal case, while the first term corresponds to $\mathbb{E}\{Z_i^2\}$, which indicates that the energy reflected from each RIS element is uniformly distributed across the $2K$ selected antennas, covering both the real and imaginary parts. This behavior is a direct result of the proposed optimization algorithm.

Therefore, an approximate ABEP upper bound for the case of optimal phase shifts can be obtained by combining \textit{Case}~$\mathcal{C}_1$ with \textit{Case}~$\mathcal{C}_3$, and \textit{Case}~$\mathcal{C}_2$ with \textit{Case}~$\mathcal{C}_4$ and using (\ref{eq:optimal_var}).

To provide further insights, Table~\ref{tab:var} reports numerical results for the variance of the noise-free signal at a selected receive antenna and compares them with the values provided by (\ref{eq:optimal_var}). The results confirm that (\ref{eq:optimal_var}) offers a sufficiently accurate approximation of the variance of the noise-free received signal. 

\begin{table}

\caption{Noise-free signal variance at a selected antenna using optimal phase shifts. Here $10^4$ channel realizations were used for numerical results. \label{tab:var}}

\centering{}%
\begin{tabular}{|c|c|c|c|}
\hline 
$N$ & $K$ & $\sigma^2$ calculated using (\ref{eq:optimal_var}) & Numerical variance \\
\hline
\multirow{2}{*}{$256$} & $2$ & $13.735$ & $13.826$ \tabularnewline
\cline{2-4} \cline{3-4} \cline{4-4}
 & $3$ & $9.156$ & $9.057$ \tabularnewline
\cline{2-4} \cline{3-4} \cline{4-4}
\hline
\multirow{2}{*}{$512$} & $2$ & $27.469$ & $27.580$ \tabularnewline
\cline{2-4} \cline{3-4} \cline{4-4}
 & $3$ & $18.313$ & $18.491$ \tabularnewline
\cline{2-4} \cline{3-4} \cline{4-4}

\hline 
\end{tabular}
\end{table}

}

\section{Multicast Communications\label{sec:Multicast-Communications}}

\begin{figure}[t]
\begin{centering}
\includegraphics[scale=0.2]{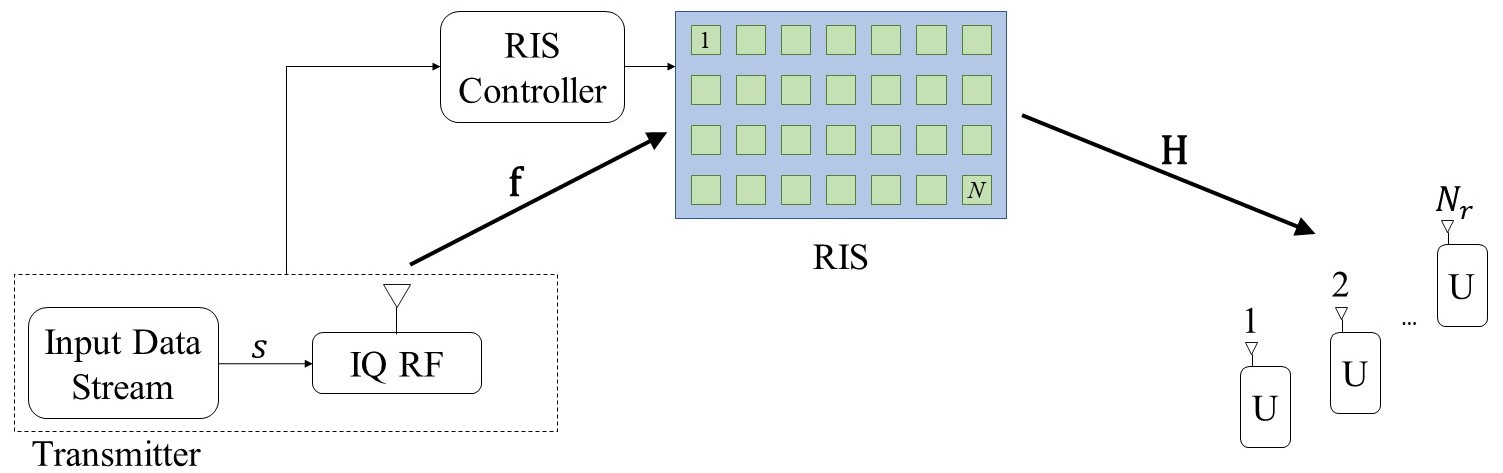}
\par\end{centering}
\caption{A diagrammatic illustration of RIS-assisted multicast communications.\label{fig:Schematic-multicast}}
\end{figure}

In this section, we consider the special case where all receive antennas
are active, i.e., no IM is used. A schematic diagram of the proposed multicast system is presented
in Fig.~\ref{fig:Schematic-multicast}. We consider a multicast channel
in which a group of $N_r$ single-antenna users receives a shared
single stream of information from the transmitter. Hence, the phase
shifts of the RIS elements are adjusted only to target the receive antennas,
and not to transmit any information bits through SM
or polarity bits. As a result, in this case the transmitter needs to change the phase shifts of the RIS elements for each channel realization, not for each symbol interval. The baseband signal received by user $l$ is given by 
\begin{equation}
y_{l}=\mathbf{h}_{l}(\boldsymbol{\theta}\odot\mathbf{f})s+n_{l},\label{eq:received signal-multicast}
\end{equation}
where $s$ is the transmit symbol with $\mathbb{E}\{|s|^{2}\}=E_{s}$.
We aim to maximize the received signal energy at \emph{all} users' receive
antennas by adjusting the phase shifts of the RIS elements, $\boldsymbol{\theta}$. It can be shown that the max-min optimization problem used to solve for the optimal vector $\boldsymbol{\theta}$, enforces the equality of the resulting optimal values. As a result, all users encounter the same SNR at their respective receive antennas. This characteristic is of particular interest in multicast communications, where equal SNR among users is a desirable aspect.
Expanding (\ref{eq:received signal-multicast}), the received signal can
be expressed as 
\begin{align*}
y_{l} & =\left[\sum_{i=1}^{N}\beta_{i}h_{l,i}e^{j(\psi_{i}+\phi_{i})}\right]s+n_{l} \triangleq G_{l}s+n_{l},
\end{align*}
where we define $G_{l}$
as the \emph{effective} coefficient of the RIS-assisted channel between
the transmit antennas and user $l$. With the purpose of simplifying
the detection procedure at the receivers, and with an interest in
equalizing the received signals, we design the phase shifts such that the values $G_l^\Re$ for $l \in [N_r]$ are maximized. Then, the optimization problem
can be defined as 
\begin{align*}
\underset{\{\omega_{i}^{\Re}\},\{\omega_{i}^{\Im}\}}{\max} & \ \min\left\{ \sum_{i=1}^{N}\beta_{i}\left(h_{l,i}^{\Re}\omega_{i}^{\Re}-h_{l,i}^{\Im}\omega_{i}^{\Im}\right):l\in[N_{r}]\right\} \\
\mbox{s.t.}\;\; & \ \left(\omega_{i}^{\Re}\right)^{2}+\left(\omega_{i}^{\Im}\right)^{2}=1,\ \forall i \in [N],
\end{align*}
where $\omega_{i}=e^{j(\psi_{i}+\phi_{i})}$. This optimization problem
is similar to (\ref{eq: max-min-QSM}), and can be solved in a similar manner.
Hence, the non-convex optimization problem involving $N$ \textit{complex} variables
is reduced to solving a nonlinear system of $(N_{r}-1)$ equations in
$(N_{r}-1)$ \textit{real} variables, where $N_{r}\ll N$. It is worth noting that
this design technique operates as a \emph{zero forcing} (ZF) \emph{pre-equalizer},
i.e., \emph{equal} noise-free signals are received at all receive
antennas. Finally, at the receiver, a one-dimensional maximum likelihood (ML) detector can be used to detect the transmitted symbols. The ML detector operates via 
\begin{align}
\hat{s} & =\arg\min_{s}\{|y_{l}-G_{l}s|\} \label{eq:ML6}\\
 & \approx \arg\min_{s}\{|y_{l}-G_{l}^{\Re}s|\} \label{eq:approximate-ML}.
\end{align}
Note that each user only needs to know the effective channel coefficient
$G_{l}^{\Re}$, which is equal for all users, as the optimization problem equalizes the resulting optimum values.

\begin{figure*}[t]
\begin{centering}
\includegraphics[scale=0.380]{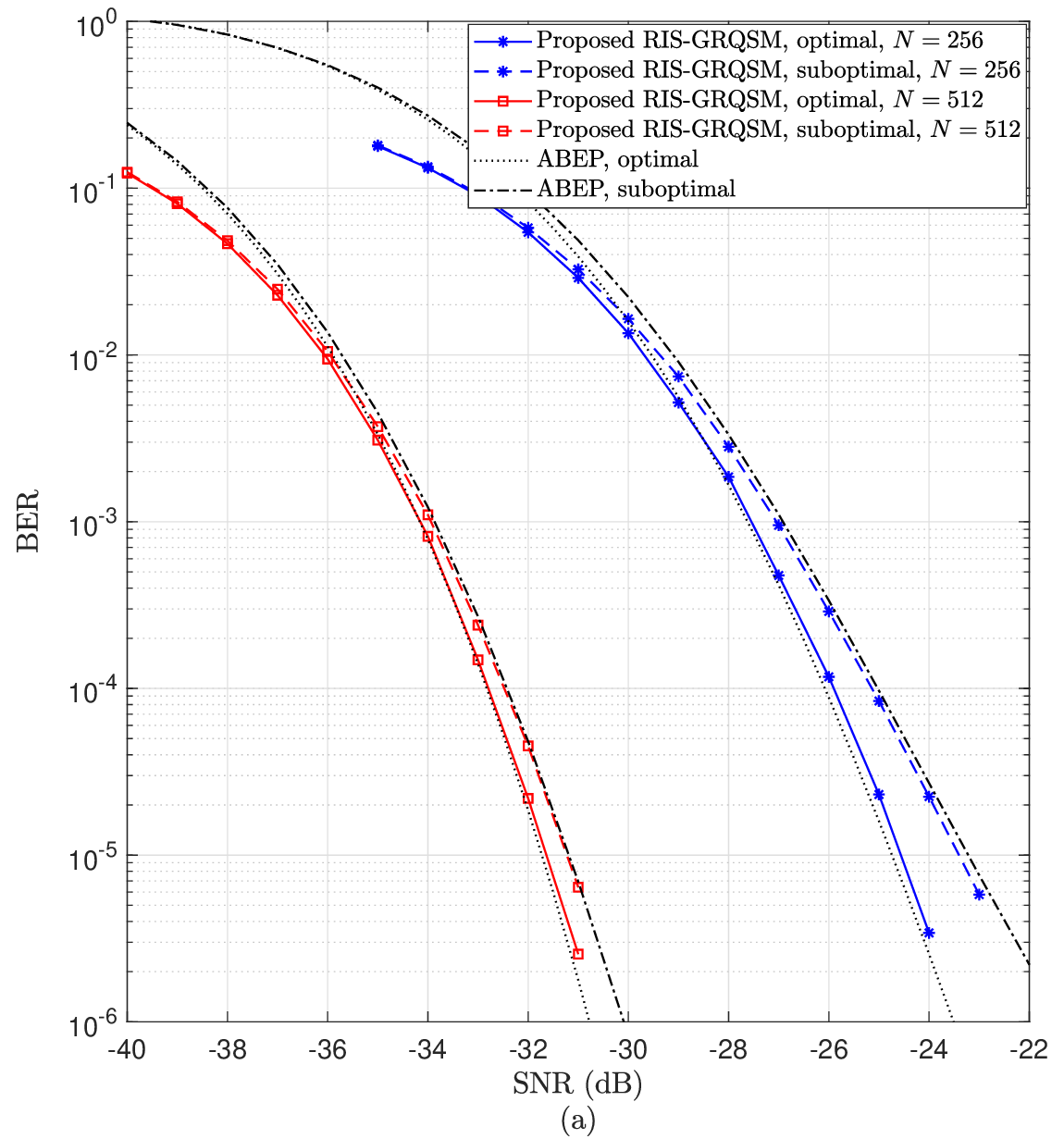}~~~~~~~~~~~~~~~~\includegraphics[scale=0.380]{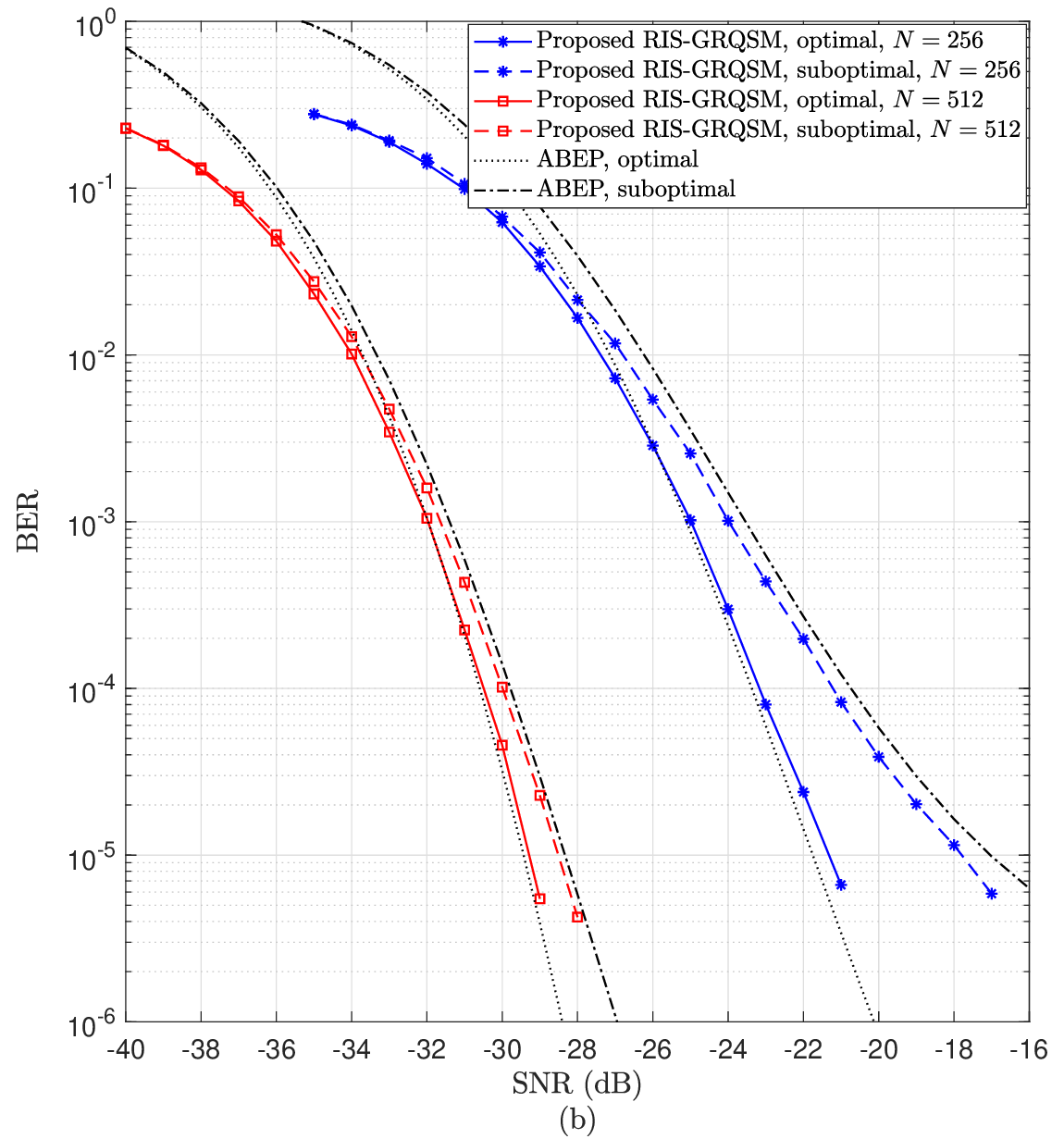}
\par\end{centering}
\caption{BER performance of the proposed RIS-GRQSM system; $N_{r}=8$,
and (a) $R=12$~bpcu ($K=2$), (b) $R=16$~bpcu ($K=3$).\label{fig:BER-performance}}
\end{figure*}

{\color{black}
\subsection*{ABEP performance of multicast communications}
In this subsection, we analyze the ABEP of the proposed multicast system. To the best of our knowledge, such an analysis has not appeared in the literature due to the complexity of the optimization algorithm needed to obtain the optimal phase shifts. Considering (\ref{eq:ML6}) and (\ref{eq:approximate-ML}), the PEP is given by 
\begin{align}
    {\rm PEP}(s\rightarrow\hat{s}|G_l)&={\rm Q}\left(\sqrt{\frac{|G_l|^2 d^2_{s,\hat{s}}E_s}{2N_0}}\right) \nonumber\\
    & \approx
    {\rm Q}\left(\sqrt{\frac{(G_l^{\Re})^2 d^2_{s,\hat{s}}E_s}{2N_0}}\right), \label{eq:PEP_multicast}
\end{align}
where $d_{s,\hat{s}}$ is the normalized Euclidean distance between the pair of symbols $(s,\hat{s})$. Hence, an upper bound on the ABEP is given by
\begin{align}
    {\rm ABEP_{ub}}=&\frac{1}{M\log_2M} \sum_s \sum_{\hat{s}} e(s,\hat{s}) \nonumber \\
    &\times \mathbb{E}\left\{{\rm Q}\left(\sqrt{\frac{({G_l^{\Re}})^2 d^2_{s,\hat{s}}E_s}{2N_0}}\right)\right\}, \label{eq:ABEP_multicast}
\end{align}
where $e(s,\hat{s})$ is the number of bit errors between the pair of symbols. The average PEP can be evaluated using (\ref{eq:craig}). To this end, taking into account (\ref{eq:optimal_var}) and Theorem~\ref{thm:mean_V}, $G_l^\Re$ can be approximated as normally distributed with mean $\frac{N\pi}{4\sqrt{N_r}}$ and variance $\frac{N}{Nr}\left(1-\frac{\pi^2}{16}\right)$.

\section{Numerical Results\label{sec:Numerical-Results6}}

In this section, we demonstrate the BER performance of the proposed systems through numerical simulations and compare the results with the most relevant benchmark schemes from the literature.
\subsection{RIS-GRQSM system}
In this subsection, we present the performance results of the proposed RIS-GRQSM system. As benchmarks, we consider the RIS-RQSM system \cite{dinan2023ris}, and a generalized version of the RIS-RQSSK system using the RIS partitioning approach \cite{zhang2022gssk}\footnote{The authors of \cite{zhang2022gssk} proposed an RIS partitioning approach (i.e., the RIS elements are partitioned into multiple sets, each of which targets one receive antenna) in order to extend the RIS-SSK system to the RIS-GSSK scheme. Our preliminary results have shown a significant advantage over this scheme; hence, as a more relevant benchmark, we have implemented a system that employs an approach similar to RIS-GSSK in order to generalize the RIS-RQSSK system.}, both of which incorporate the concept of SM at the receiver.

\begin{figure*}[t]
\begin{centering}
\includegraphics[scale=0.38]{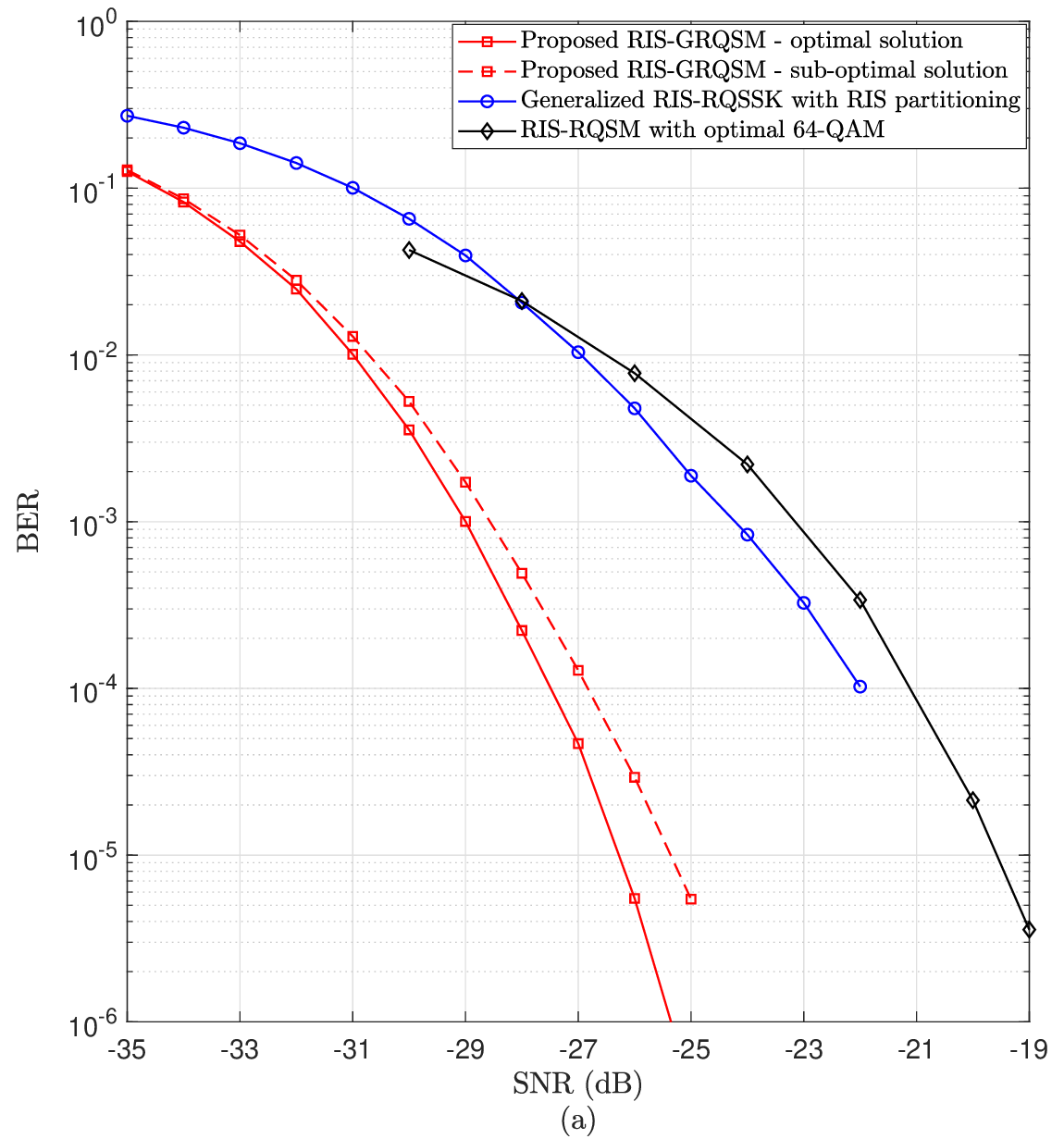}~~~~~~~~~~~~~~~~\includegraphics[scale=0.38]{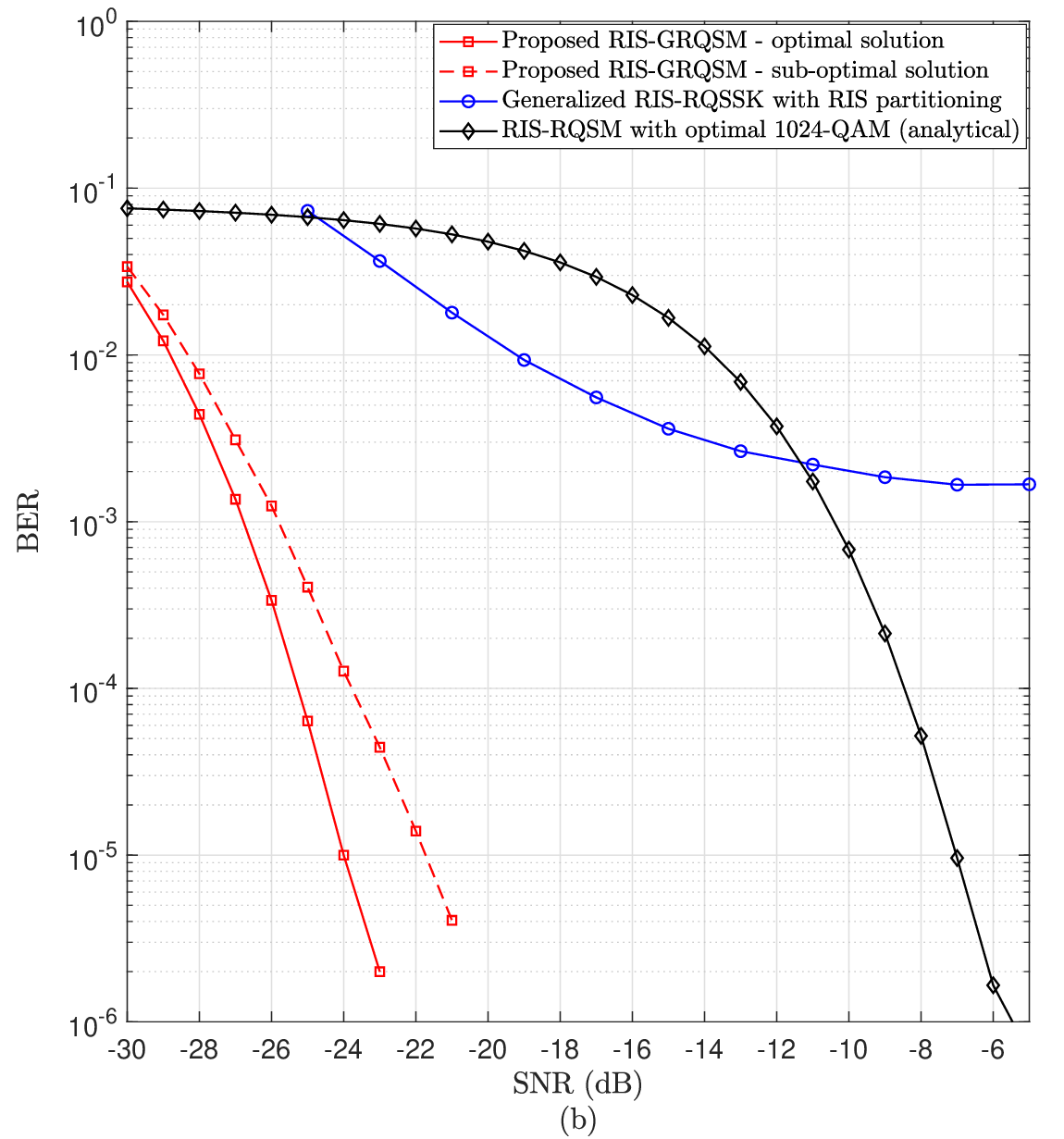}
\par\end{centering}
\caption{Comparison of the BER performance of the proposed RIS-GRQSM system
with that of RIS-RQSM and generalized RIS-RQSSK systems, for $N=256$, $N_{r}=8$,
and (a) $R=12$~bpcu ($K=2$), (b) $R=16$~bpcu ($K=3$).\label{fig:BER-performance1}}
\end{figure*}

In Fig.~\ref{fig:BER-performance}, the BER performance of the proposed RIS-GRQSM system is evaluated for different values of $N$ and $K$, with a fixed number of receive antennas $N_{r}=8$. Fig.~\ref{fig:BER-performance}(a) shows results for $K=2$ selected receive antennas, corresponding to a SE of $R=12$~bpcu, while Fig.~\ref{fig:BER-performance}(b) presents results for $K=3$, yielding $R=16$~bpcu. For each case, results are provided for two different numbers of RIS elements: 256 and 512.

It is evident that the proposed system achieves strong performance even at lower SNRs, due to the signal enhancement provided by the RIS. Furthermore, it can be observed that doubling the number of RIS elements from 256 to 512 significantly enhances the BER performance, achieving an improvement of more than 6 dB. This gain aligns with observations for existing RIS-aided communication systems, where increasing the number of reflecting elements significantly improves the signal quality \cite{basar2019wireless,dinan2022ris,dinan2023ris}.

Fig.~\ref{fig:BER-performance} also shows the BER performance of the RIS-GRQSM system where the closed-form suboptimal solution for the RIS phase shifts is adopted. It can be observed that, as expected, the suboptimal approach performs close to the optimal one when the ratio $N/(2K)$ is sufficiently large. However, as $K$ increases, the performance gap between the suboptimal and optimal methods is more significant. Moreover, the analytical ABEP curves are also plotted, which verify the accuracy of the theoretical analysis\footnote{It is worth mentioning that, in order to present more accurate analytical curves, the exact values of $\rho$ and the probabilities $P_1$, $P_2$, $P_3$, $P_4$ were computed for each simulation scenario.}.

Fig.~\ref{fig:BER-performance1} compares the BER performance of the proposed RIS-GRQSM system against that of prominent benchmark schemes. To ensure a fair comparison, the RIS-RQSM system benchmark employs an optimized modulation scheme as proposed in \cite{dinan2023ris}. Since this scheme assumes a deterministic channel between the transmitter and the RIS, we adopt the same model by setting $f_i = 1$ for all $i \in [N]$. As shown in Fig.~\ref{fig:BER-performance1}(a), the proposed scheme significantly outperforms both benchmarks. Specifically, RIS-GRQSM achieves approximately 5.5~dB and 7~dB gains over the RIS-RQSM and generalized RIS-RQSSK systems, respectively, at the BER of $10^{-5}$. In Fig.~\ref{fig:BER-performance1}(b), with $K=3$ ($R=16$~bpcu), the RIS-RQSM system requires a 1024-QAM constellation to match the throughput. In this setting, the performance advantage of the proposed RIS-GRQSM scheme is even more substantial.

This performance superiority stems from two key factors: (i) the RIS-RQSM system relies on the use of higher-order modulation, which leads to reduced Euclidean distances between symbols and thus to performance degradation; and (ii) the generalized RIS-RQSSK scheme suffers from a reduced allocation of RIS elements per active antenna, causing an error floor. Although the RIS-GRQSM approach introduces additional transmitter complexity due to the optimization requirement, the proposed suboptimal solution still delivers substantial performance gains. These results highlight both the effectiveness and practical viability of the RIS-GRQSM system compared to existing alternatives.
}

\begin{figure*}[t]
\begin{centering}
\includegraphics[scale=0.380]{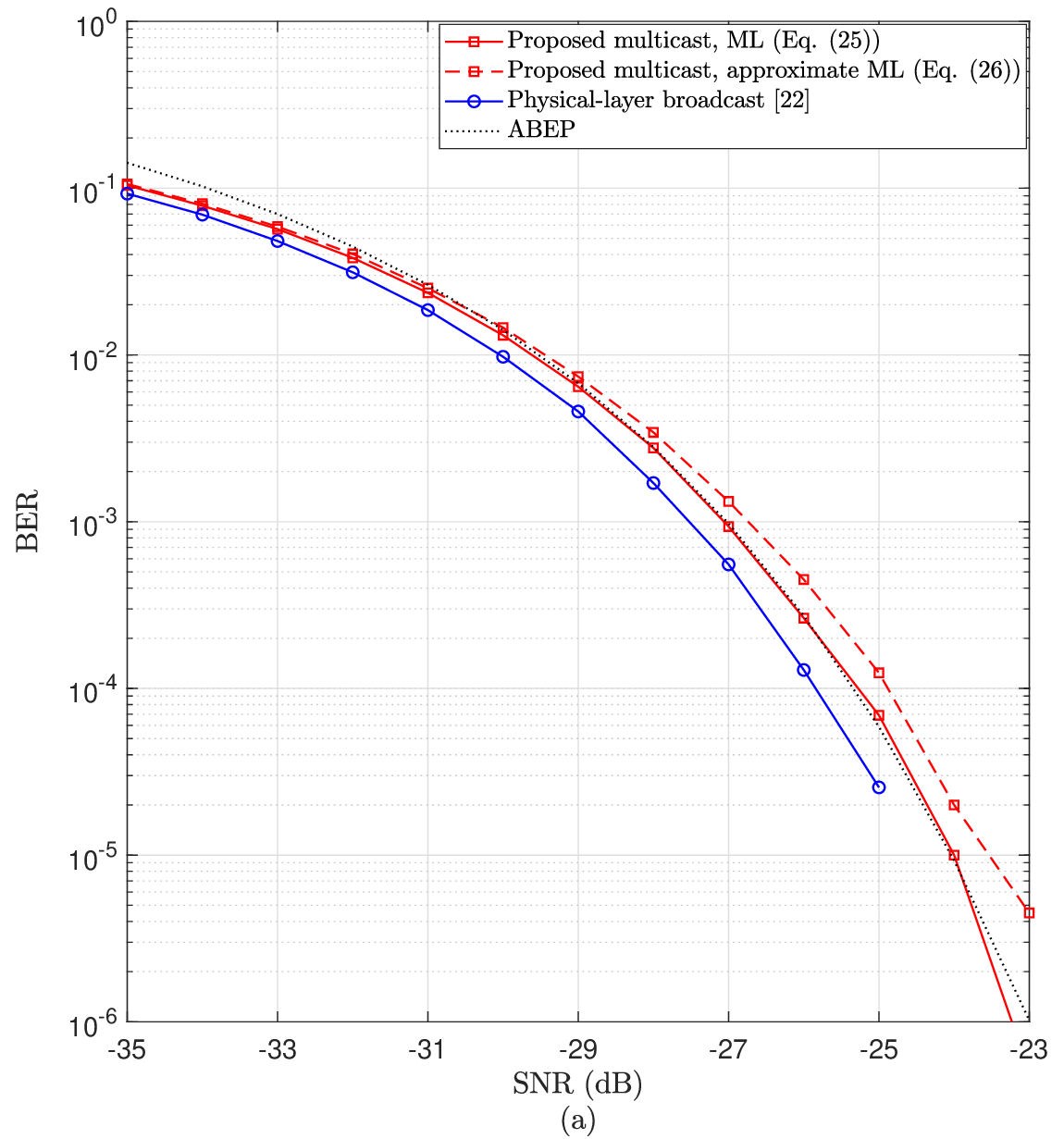}~~~~~~~~~~~~~~~~\includegraphics[scale=0.380]{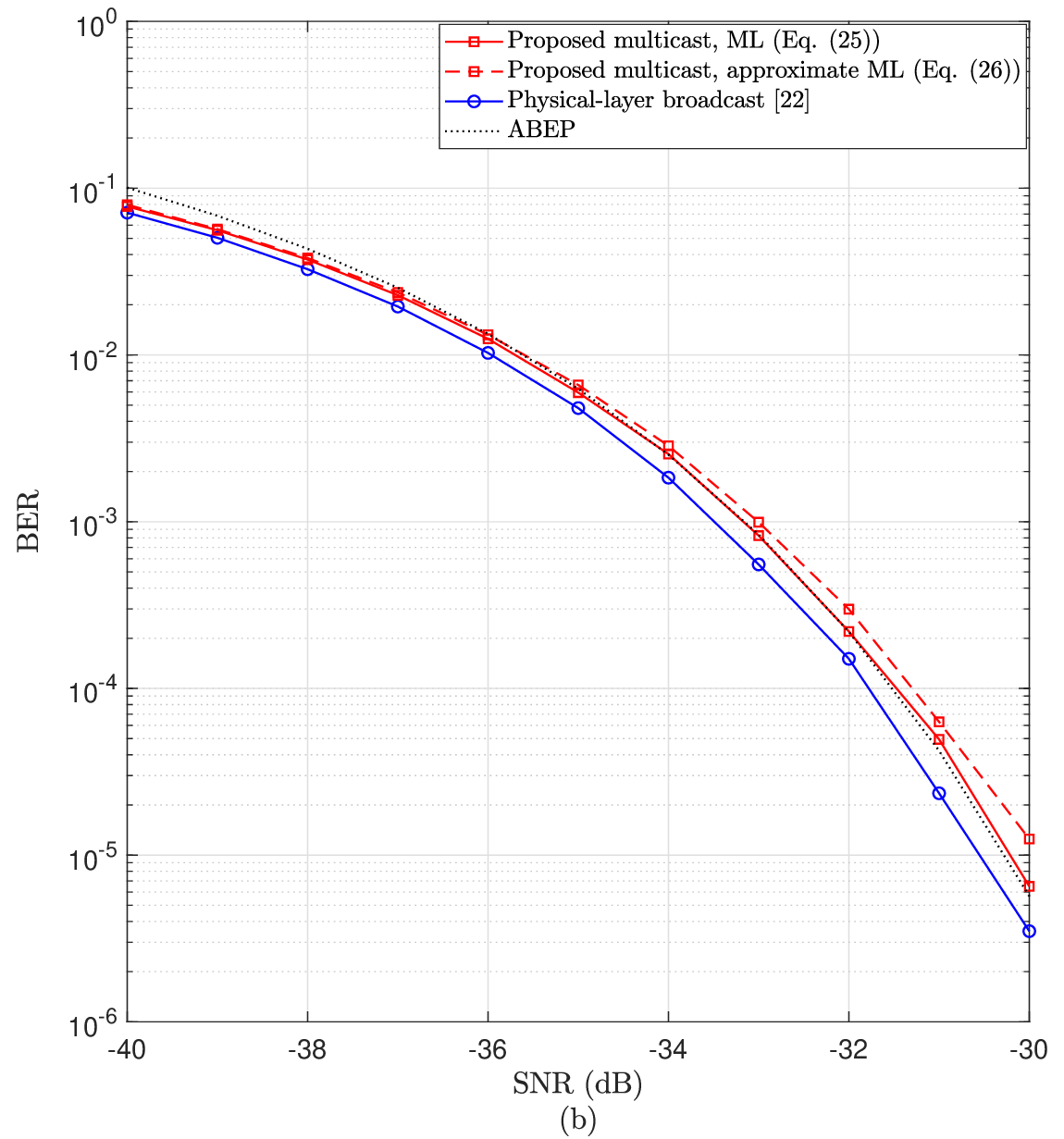}
\par\end{centering}
\caption{Comparison of the BER performance of the proposed multicast system
with that of the physical-layer broadcast scheme proposed in \cite{han2020broadcast}, where 4-QAM is used, $N_{r}=2$, $N_{t}=1$
and (a) $N=128$, (b) $N=256$.\label{fig:BER-multicast}}
\end{figure*}

\subsection{Multicast Communications}
Fig.~\ref{fig:BER-multicast} presents the BER performance of the proposed scheme in the case of multicast communications, where 4-QAM is used for IQ modulation in systems with $N=128$ and $N=256$. In addition, we compare the BER results with that of the benchmark scheme which is the physical-layer broadcast system proposed in \cite{han2020broadcast}. For the benchmark scheme, we employ the semi-definite relaxation (SDR) algorithm proposed in \cite{han2020broadcast} to optimize the phase shifts of the RIS elements where $N_t=1$, which has a significantly high complexity that makes it practically infeasible to implement (it was shown in \cite{han2020broadcast} that this complexity is $\mathcal{O} ( N^7 )$). In contrast, the proposed system uses the solution provided in (\ref{eq:sys_eq_1})-(\ref{eq:sys_eq_3}) which involves a system of $N_r-1$ nonlinear equations where $N_r\ll N$. 
Despite the fact that the proposed solution has significantly lower complexity compared to the SDR method presented in \cite{han2020broadcast}, the proposed system achieves a BER performance that closely matches that of the benchmark. The primary advantage of the proposed solution lies in its notable complexity reduction, which makes it a practical and feasible approach. The performance gap diminishes as the number of RIS elements increases, as with a larger number of RIS elements the system exhibits improved capability in performing ZF equalization. In Fig.~\ref{fig:BER-multicast}, we also illustrate the performance of the approximate ML detector given by (\ref{eq:approximate-ML}). It is observed that the approximate ML detector performs nearly as well as the ML detector, confirming that the proposed solution functions similarly to the ZF pre-equalizer in conventional multicast MIMO communications. We also present the theoretical ABEP curves that agree closely with the simulation results, thus validating the theoretical analysis.

\begin{figure}[t]
\begin{centering}
\includegraphics[scale=0.40]{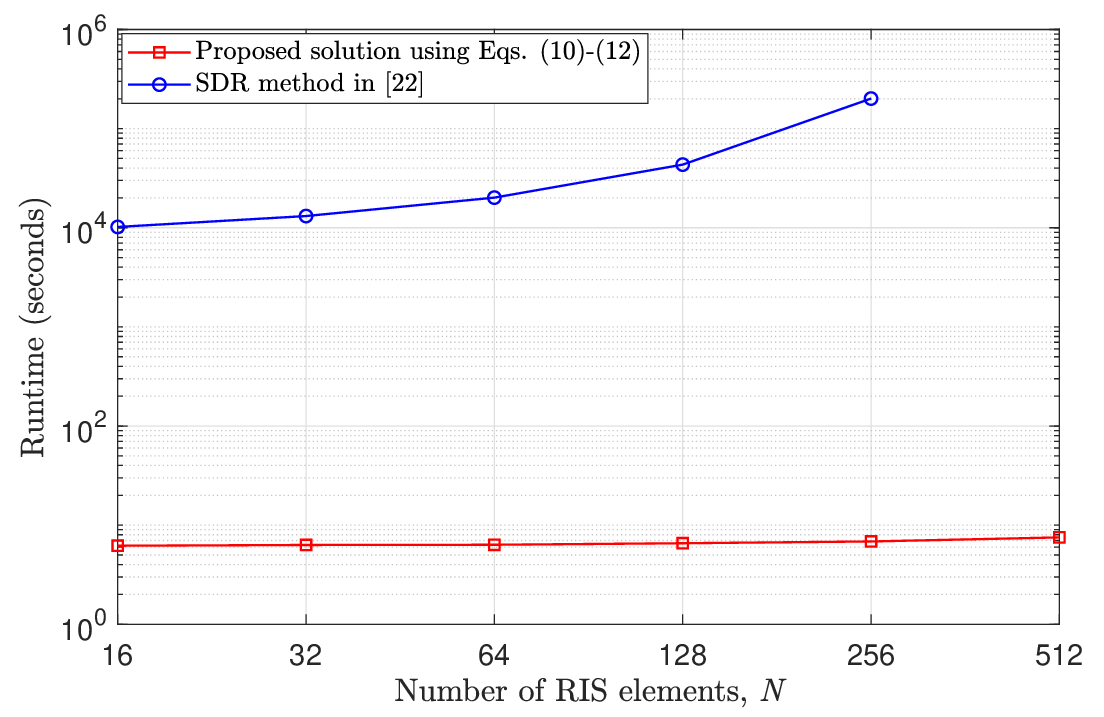}
\par\end{centering}
\caption{Runtime comparison of the proposed solution for optimizing the phase shifts of the RIS elements using (\ref{eq:sys_eq_1})-(\ref{eq:sys_eq_3})
with that of the SDR method proposed in \cite{han2020broadcast}. Here, $N_r=2$ and $10^4$ channel realizations are used.\label{fig:runtime}}
\end{figure}

Finally, Fig.~\ref{fig:runtime} illustrates the runtime comparison of the solution provided in (\ref{eq:sys_eq_1})-(\ref{eq:sys_eq_3}) with that of the benchmark, i.e., the SDR method. It can be observed that in the SDR method, the time spent to optimize the phase shifts of the RIS elements is significantly higher than that of the proposed solution. In addition, the runtime of the proposed solution is approximately constant with respect to the number of RIS elements, as the complexity mainly depends on the number of nonlinear equations (which in turn depends only on the number of users). In contrast, the runtime of the SDR method dramatically increases with an increasing number of RIS elements.

\section{Conclusion\label{sec:Conclusion6}}

In this paper, we have proposed a novel RIS-assisted index modulation scheme, termed RIS-assisted generalized receive quadrature spatial modulation (RIS-GRQSM), which extends existing RIS-assisted SM frameworks by activating multiple receive antennas independently for the in-phase and quadrature components. A max–min optimization problem was formulated to adjust the RIS phase shifts, and by exploiting Lagrange duality, the original non-convex optimization was reduced to a system involving a much smaller set of real-valued variables. Furthermore, a closed-form suboptimal solution was developed, offering near-optimal performance when the RIS has a sufficiently large number of reflecting elements. To simplify receiver design, we also introduced a low-complexity non-coherent energy-based greedy detector. In addition, we extended the RIS-GRQSM framework to a multicast communication scenario, where all users receive identical information with equal SNR levels. We analyzed the average bit error probability (ABEP) of the proposed systems under both optimal and suboptimal optimization strategies, which is rarely feasible in RIS-aided communication studies due to the inherent complexity of phase-shift optimization. Simulation results confirmed that RIS-GRQSM significantly improves spectral efficiency and error rate performance compared to benchmark schemes. For the multicast case, analytical and numerical results demonstrated that the proposed solution achieves a favorable trade-off between BER performance and computational complexity.

Overall, the proposed RIS-GRQSM and its extension to multicast communication provide energy-efficient, spectrally efficient, and low-complexity transceiver designs, making them promising candidates for enabling high-capacity wireless networks in future beyond-5G (B5G) systems.

\appendices{}
{\color{black}
\section{Proof of Theorem \ref{thm:mean_V}\label{app:appendix1}}
In this appendix, we analyze the expected value and variance of $Z_i$ and $\hat{Z}_i$ for \textit{Case}~$\mathcal{C}_1$, which are then used to evaluate the corresponding values for $Z$ and $\hat{Z}$. For the sake of brevity, we omit the proof for \textit{Cases~}$\mathcal{C}_2$, $\mathcal{C}_3$ and $\mathcal{C}_4$ which can be derived in a similar manner.  In the following, we simplify notation by omitting the index $i$ in the expressions for $Z_i$ and $\hat{Z}_i$.

It is worth recalling that, in \textit{Case}~$\mathcal{C}_1$, $m_k,m_q\notin \mathcal{N}$. Here, $k$ denotes the index of the desired antenna among the selected antennas used for calculating the mean and variance.

\subsection{Expected value of $Z_i$}
The average of $Z_i$, considering the symmetry between variables, is given by 
\begin{equation}
    \mathbb{E}\{Z_i\}=\mathbb{E}\{\beta\} \left(2\mathbb{E}\left\{\frac{A_k^2}{\sqrt{W}}\right\}+(4K-2)\mathbb{E}\left\{\frac{A_kA_j}{\sqrt{W}}\right\}\right), \label{eq:E_Z_1}
\end{equation}
where $j\in [K]$ and $j\neq k$,
$W\triangleq\big(\sum_{j'=1}^{K}A_{j'}+\sum_{j'=1}^{K}B_{j'}\big)^{2}+\big(\sum_{j'=1}^{K}C_{j'}+\sum_{j'=1}^{K}D_{j'}\big)^{2}$, and $\mathbb{E}\{\beta\}=\sqrt{\pi}/2$.

To find the expected value, we analyze each term in (\ref{eq:E_Z_1}) separately.

For the first term, according to the law of total expectation (LTE), we can write  
\begin{equation}
    \mathbb{E}\left\{\frac{A_k^2}{\sqrt{W}}\right\}=\mathbb{E}_{A_k}\left\{A_k^2\mathbb{E}_{W|A_k}\left\{W^{-\frac{1}{2}}|A_k\right\}\right\}, \label{eq:A2_sqW}
\end{equation}
where the right-hand side (RHS) requires the calculation of the inverse and fractional moment of $(W|A_k)$, which is given by \cite{mathai1992quadratic} 
\begin{equation}
    \mathbb{E}_{W|A_k}\left\{W^{-p}|A_k\right\}=\frac{1}{\Gamma(p)}\int_0^\infty s^{p-1} \mathfrak{L}_s(f_W(W|A_k)){\rm d}s, \label{eq:E_W|A}
\end{equation}
where $f_W(W|A_k)$ is the PDF of $W$ conditioned on $A_k$, and $\mathfrak{L}_s$ denotes its Laplace transform (LT) which can be expressed as 
\begin{align}
   \mathfrak{L}_s(f_W(W|A_k))= & \Big(\frac{1}{1+2Ks}\Big)^{\frac{1}{2}} \Big(\frac{1}{1+(2K-1)s}\Big)^{\frac{1}{2}} \nonumber \\ 
   & \times \exp\Big(\frac{-A_k^2s}{1+(2K-1)s}\Big). \label{eq:laplace1}
\end{align}
Hence, considering $A_k\sim\mathcal{N}(0,1/2)$, we may substitute (\ref{eq:E_W|A}) into (\ref{eq:A2_sqW}) to obtain
\begin{align}
    & \mathbb{E}\left\{\frac{A_k^2}{\sqrt{W}}\right\}= \nonumber \\
    & \frac{1}{\pi}\int_0^\infty s^{-\frac{1}{2}} \Big(\frac{1}{1+2Ks}\Big)^{\frac{1}{2}} \Big(\frac{1}{1+(2K-1)s}\Big)^{\frac{1}{2}} \nonumber\\
    & \times \left(\int_{-\infty}^\infty A_k^2 \exp\Big(-A_k^2\frac{1+2Ks}{1+(2K-1)s}\Big){\rm d}A_k\right){\rm d}s, \nonumber
\end{align}
which after some manipulations results in
\begin{equation}
    \mathbb{E}\left\{\frac{A_k^2}{\sqrt{W}}\right\}=\frac{(4K-1)\sqrt{\pi}}{8K\sqrt{2K}}.
\end{equation}
We can follow the same procedure for the second term using the LTE. Hence, considering $A_k$ and $A_j$ are constants, the LT is given by 
\begin{align}
   \mathfrak{L}_s(f_W(W|A_k,A_j))= & \Big(\frac{1}{1+2Ks}\Big)^{\frac{1}{2}} \Big(\frac{1}{1+(2K-2)s}\Big)^{\frac{1}{2}} \nonumber \\ 
   & \times \exp\Big(\frac{-(A_k+A_j)^2s}{1+(2K-2)s}\Big). \label{eq:laplace2}
\end{align}
Next, utilizing the PDFs of $A_k$ and $A_j$, we can write 
\begin{align}
     \mathbb{E}\left\{\frac{A_kA_j}{\sqrt{W}}\right\}= & \frac{1}{\sqrt{\pi}}\int_0^\infty s^{-\frac{1}{2}} \Big(\frac{1}{1+2Ks}\Big)^{\frac{1}{2}} \Big(\frac{1}{1+(2K-2)s}\Big)^{\frac{1}{2}} \nonumber\\
    & \times \bigg(\iint_{-\infty}^\infty \frac{1}{\pi} A_kA_j \exp\Big(\frac{-(A_k+A_j)^2s}{1+(2K-2)s} \nonumber \\
    & -A_k^2-A_j^2\Big){\rm d}A_k{\rm d}A_j\bigg){\rm d}s. \nonumber
\end{align}
We note that the exponential term within the inner double integral is in the form of a bivariate normal distribution.
Hence, we can rewrite the inner double integral as
\begin{align}
    \frac{(1+(2K-2)s)^{1/2}}{(1+2Ks)^{1/2}}\iint_{-\infty}^\infty \frac{A_kA_j}{\sqrt{(2\pi)^2|\boldsymbol{\Sigma}_1|}} & \nonumber\\
    \times \exp\Big({-\frac{1}{2}[A_k,A_j]\boldsymbol{\Sigma}_1^{-1}[A_k,A_j]^{\rm T}}\Big){\rm d}A_k{\rm d}A_j, & \label{eq:multiNorm}
\end{align}
where 
\begin{equation}
    \boldsymbol{\Sigma}_1=\frac{1}{2(1+2Ks)}
    \begin{bmatrix}
        1+(2K-1)s & -s \\
        -s        & 1+(2K-1)s
    \end{bmatrix}
\end{equation}
is the covariance matrix. Therefore, the double integral in (\ref{eq:multiNorm}) represents the covariance between the two variables, which is equal to $\frac{-s}{2(1+2Ks)}$. Consequently, we obtain 
\begin{equation}
    \mathbb{E}\left\{\frac{A_kA_j}{\sqrt{W}}\right\}= \frac{-1}{2\sqrt{\pi}}\int_0^\infty \frac{s^{1/2}}{(1+2Ks)^2}{\rm d}s=\frac{-\sqrt{\pi}}{8K\sqrt{2K}}.
\end{equation}
Finally, the expected value of $Z_i$ is given by
\begin{equation}
    \mathbb{E}\{Z_i\}=\frac{\sqrt{\pi}}{2} \Big(\frac{2(4K-1)\sqrt{\pi}}{8K\sqrt{2K}}-\frac{(4K-2)\sqrt{\pi}}{8K\sqrt{2K}}\Big)=\frac{\pi}{4\sqrt{2K}}. \label{eq:E_Z1_final}
\end{equation}

\subsection{Variance of $Z_i$}
The variance can be calculated as $\mathbb{V}\{Z_i\}=\mathbb{E}\{Z_i^2\}-\mathbb{E}^2\{Z_i\}$. Taking the symmetry between the variables, we can write 
\begin{align}
    \mathbb{E}\{Z_i^2\}= & 2\mathbb{E}\Big\{\frac{A_k^4}{W}\Big\}+4(2K-1)\mathbb{E}\Big\{\frac{A_k^3A_j}{W}\Big\}+2(2K-1) \nonumber\\
    & \times\mathbb{E}\Big\{\frac{A_k^2A_j^2}{W}\Big\}+2(2K-1)(2K-2)\mathbb{E}\Big\{\frac{A_k^2A_jA_{j'}}{W}\Big\} \nonumber\\
    & +2\mathbb{E}\Big\{\frac{A_k^2C_k^2}{W}\Big\}+4(2K-1)\mathbb{E}\Big\{\frac{A_k^2C_kC_j}{W}\Big\} \nonumber\\
    & +2(2K-1)^2\mathbb{E}\Big\{\frac{A_kC_kA_jC_j}{W}\Big\}, \label{eq:E_Z2}
\end{align}
where $j,j'\in [K]\ \&\ j\neq j'\neq k$. Note that $\mathbb{E}\{\beta^2\}=1$. For brevity, we only present the calculation of the last term, i.e., $\mathbb{E}\Big\{\frac{A_kC_kA_jC_j}{W}\Big\}$. Other terms can be calculated using the same procedure. 

Using LTE, considering (\ref{eq:E_W|A}) and given $(A_k,A_j,C_k,C_j)$, the expected value in the last term is given by 
\begin{align}
    & \mathbb{E}\Big\{\frac{A_kC_kA_jC_j}{W}\Big\}= \nonumber\\
    & \int_0^\infty \mathbb{E}\big\{ A_kC_kA_jC_j\mathfrak{L}_s(f_W(W|A_k,A_j,C_k,C_j))\big\}{\rm d}s. \label{eq:E_var}
\end{align}
The LT of $W$ is expressed as 
\begin{align}
    & \mathfrak{L}_s(f_W(W|A_k,A_j,C_k,C_j))= \nonumber \\ 
    & \frac{1}{1+(2K-2)s} \exp\Big(\frac{-((A_k+A_j)^2+(C_k+C_j)^2)s}{1+(2K-2)s}\Big). \label{eq:laplace3}
\end{align}
Next, similarly to (\ref{eq:multiNorm}), the expected value in the RHS of (\ref{eq:E_var}) is given by 
\begin{align}
    & \mathbb{E}\big\{ A_kC_kA_jC_j\mathfrak{L}_s(f_W(W|A_k,A_j,C_k,C_j))\big\}= \nonumber\\
    & \frac{1}{1+2Ks}\iiiint_{-\infty}^\infty \frac{Q_1Q_2}{\sqrt{(2\pi)^4|\boldsymbol{\Sigma}_2|}} \exp\Big({-\frac{1}{2}\mathbf{u}^{\rm T}\boldsymbol{\Sigma}_2^{-1}\mathbf{u}}\Big){\rm d}\mathbf{u}, & \label{eq:multiNorm4}
\end{align}
where 
\begin{equation}
    \boldsymbol{\Sigma}_2=
    \begin{bmatrix}
        \boldsymbol{\Sigma}_1 & \boldsymbol{0} \\
        \boldsymbol{0}        & \boldsymbol{\Sigma}_1
    \end{bmatrix},
\end{equation}
$\mathbf{u}\triangleq [A_k,A_j,C_k,C_j]^{\rm T}$, and $Q_1$ and $Q_2$ are quadratic forms defined as $Q_1\triangleq \mathbf{u}^{\rm T}\boldsymbol{\Delta}_1\mathbf{u}$ and $Q_2\triangleq \mathbf{u}^{\rm T}\boldsymbol{\Delta}_2\mathbf{u}$, where 
\begin{equation}
    \boldsymbol{\Delta}_1=
    \begin{bmatrix}
        0 & 1/2 & 0 & 0 \\
        1/2 & 0 & 0 & 0 \\
        0 & 0 & 0 & 0 \\
        0 & 0 & 0 & 0
    \end{bmatrix},\ 
    \boldsymbol{\Delta}_2=
    \begin{bmatrix}
        0 & 0 & 0 & 0 \\
        0 & 0 & 0 & 0 \\
        0 & 0 & 0 & 1/2 \\
        0 & 0 & 1/2 & 0
    \end{bmatrix}.
\end{equation}
Hence, (\ref{eq:E_var}) is derived as \cite[Theorem~3.2d.4]{mathai1992quadratic} 
\begin{align}
    & \mathbb{E}\Big\{\frac{A_kC_kA_jC_j}{W}\Big\}= \nonumber\\
    & \int_0^\infty \frac{1}{1+2Ks}\Big(2{\rm tr}(\boldsymbol{\Delta}_1\boldsymbol{\Sigma}_2\boldsymbol{\Delta}_2\boldsymbol{\Sigma}_2)+{\rm tr}(\boldsymbol{\Delta}_1\boldsymbol{\Sigma}_2){\rm tr}(\boldsymbol{\Delta}_2\boldsymbol{\Sigma}_2)\Big){\rm d}s \nonumber\\
    & =\int_0^\infty \frac{s^2}{4(1+2Ks)^3}{\rm d}s. \label{eq:var1}
\end{align}
By employing the same procedure, the other terms in (\ref{eq:E_Z2}) are obtained as 
\begin{align}
    & \mathbb{E}\Big\{\frac{A_k^4}{W}\Big\}=\int_0^\infty \frac{3(1+(2K-1)s)^2}{4(1+2Ks)^3}{\rm d}s \\
    & \mathbb{E}\Big\{\frac{A_k^3A_j}{W}\Big\}=\int_0^\infty \frac{-3s(1+(2K-1)s)}{4(1+2Ks)^3}{\rm d}s \\
    & \mathbb{E}\Big\{\frac{A_k^2A_j^2}{W}\Big\}=\int_0^\infty \frac{2s^2+(1+(2K-1)s)^2}{4(1+2Ks)^3}{\rm d}s \\
    & \mathbb{E}\Big\{\frac{A_k^2A_jA_{j'}}{W}\Big\}=\int_0^\infty \frac{-s(1+(2K-3)s)}{4(1+2Ks)^3}{\rm d}s \\
    & \mathbb{E}\Big\{\frac{A_k^2C_k^2}{W}\Big\}=\int_0^\infty \frac{(1+(2K-1)s)^2}{4(1+2Ks)^3}{\rm d}s \\
    & \mathbb{E}\Big\{\frac{A_k^2C_kC_j}{W}\Big\}=\int_0^\infty \frac{-s(1+(2K-1)s)}{4(1+2Ks)^3}{\rm d}s. \label{eq:var7}
\end{align}
Substituting (\ref{eq:var1})-(\ref{eq:var7}) into (\ref{eq:E_Z2}), $\mathbb{E}\{Z_i^2\}$ can be expressed as 
\begin{equation}
    \mathbb{E}\{Z_i^2\}=\int_0^\infty \frac{(4K^2-2K)s+2K+3}{2(1+2Ks)^3}{\rm d}s= \frac{1}{2}+\frac{1}{4K}. 
\end{equation}
This results in 
\begin{equation}
    \mathbb{V}\{Z_i\}= \frac{1}{2}+\frac{8-\pi^2}{32K}. \label{eq:V_Z1_final}
\end{equation}

Considering (\ref{eq:E_Z1_final}) and (\ref{eq:V_Z1_final}), and using the CLT, yields the mean and variance of $Z$ as $\mu_1$ and $\sigma_1^2$ as presented in Theorem~\ref{thm:mean_V}.

\subsection{Expected value and variance of $\hat{Z}_i$}
It is easy to see that $\mathbb{E}\{\hat{Z}_i\}=0$ and $\mathbb{V}\{\hat{Z}_i\}=1/2$.

}

\bibliographystyle{ieeetr}
\bibliography{ref}

\end{document}